# Magnetic structure determination of rare-earth based, high moment, atomic laminates; potential parent materials for 2D magnets.

D. Potashnikov,[a,b] E. N. Caspi,[c] A. Pesach,[c] Q. Tao,[d] J. Rosen,[d] D. Sheptyakov,[e] H.A. Evans,[f] C. Ritter,[g] Z. Salman,[h] P. Bonfa,[i] T. Ouisse,[j] M. Barbier,[j] O. Rivin[c] and A. Keren[a]

[a] Faculty of Physics, Technion - Israeli Institute of Technology, Haifa 32000, Israel.

[b] Israel Atomic Energy Commission, P.O. Box 7061, Tel-Aviv 61070, Israel.

[c] Department of Physics, Nuclear Research Centre-Negev, P.O. Box 9001, Beer Sheva 84190, Israel.

[d] Materials Design, Department of Physics, Chemistry, and Biology (IFM), Linköping University, SE-581 83 Linköping, Sweden.

[e] Laboratory for Neutron Scattering and Imaging, Paul Scherrer Institute, CH-5232 Villigen PSI, Switzerland.

[f] Center for Neutron Research, National Institute of Standards and Technology, Gaithersburg, MD, USA.

[g] Institut Laue-Langevin, 71 Avenue des Martyrs, 38000 Grenoble, France.

[h] Laboratory for Muon Spin Spectroscopy, Paul Scherrer Institute, CH-5232 Villigen PSI, Switzerland.

[i] Department of Mathematical, Physical and Computer Sciences, University of Parma, 43124 Parma, Italy.

[j] Univ. Grenoble Alpes, CNRS, Grenoble INP, LMGP, F-38000 Grenoble, France.

## Abstract

We report muon spin rotation ($\mu$SR) and neutron diffraction on the rare-earth based magnets $(Mo_{2/3}RE_{1/3})_2AlC$, also predicted as parent materials for 2D derivatives, where RE = Nd, Gd (only $\mu$SR), Tb, Dy, Ho and Er. By crossing information between the two techniques, we determine the magnetic moment ($m$), structure, and dynamic properties of all compounds. We find that only for RE = Nd and Gd the moments are frozen on a microsecond time scale. Out of these two, the most promising compound for a potential 2D high $m$ magnet is the Gd variant, since the parent crystals are pristine with $m = 6.5 \pm 0.5\ \mu_B$, Néel temperature of $29 \pm 1$ K, and the magnetic anisotropy between in and out of plane coupling is smaller than $10^{-8}$. This result suggests that magnetic ordering in the Gd variant is dominated by in-plane magnetic interactions and should therefore remain stable when exfoliated into 2D sheets.

# I.  Introduction

Two-dimensional (2D) heterostructures based on Van der Waals layers require a variety of building blocks including, optionally, magnetic sheets. Up to date, only a handful of transition metal based 2D magnets have been found[1–4] and the magnetic moment $m$ of their parent compounds is limited to $4\,\mu_B$. Naturally, there is a need for more building blocks. The recently discovered series of crystalline, rare-earth (RE) based compounds,[5] $(Mo_{2/3}RE_{1/3})_2AlC$ $i$-MAX phases, may provide a range of 2D derivatives to the heterostructure toolbox. Selected $i$-MAX phases, of the general formula $(M'_{2/3}M''_{1/3})_2AlC$, have previously been shown to realize 2D materials by chemical exfoliation.[6] Importantly, through the concept of "targeted etching", the 2D sheets can be tailored to retain one or both of the $M$ elements,[6] suggesting RE-based 2D materials from the 3D atomic laminates investigated herein. Furthermore, theoretical predictions[7] and preliminary experimental results (not shown) suggest derivation of magnetic 2D sheets also from mechanical exfoliation.[8]

However, not much is known on the magnetic structure of the parent compounds, let alone on the potential 2D sheets. Temperature and field dependent magnetization and heat capacity measurements[5] on $(Mo_{2/3}RE_{1/3})_2AlC$ have shown that most compounds are magnetically ordered with transition temperatures ranging from $3.6\,K$ for RE = Er up to $28\,K$ for RE = Tb. Curie-Weiss plots of inverse susceptibility versus temperature gave negative Weiss temperatures suggesting that the magnetic coupling is antiferromagnetic (AFM). Neutron powder diffraction (NPD) on RE = Tb and Er found that both compounds order as spin density waves (SDW), confirming the AFM assumption at least in these two compounds.

In this work the magnetic ground states of bulk powder $(Mo_{2/3}RE_{1/3})_2AlC$ with RE = Nd, Gd, Tb, Dy, Ho, and Er are explored using muon spin rotation ($\mu$SR). The Gd variant is also explored in a single crystal form, as a function of initail muon polarizaiton direction. In addition, the series of powders is examined with NPD, excluding RE = Gd since it is a strong neutron absorber. From NPD, it is found that, like RE = Tb and Er in previous work,[5] all compounds except RE = Gd develop a SDW magnetic ordering with their propagation vector directed along the unique axis of the monoclinic unit cell. Using density functional theory (DFT) based muon site calculations, symmetry analysis, and information from the NPD data on RE = Nd, Tb, Ho, and Er, the magnetic structure of RE = Gd is also determined to be a simple AFM with an extremely small anisotropy ratio between in and out of plane coupling. Moreover, only RE = Nd and Gd exhibit long-range static magnetic order within the $\mu$SR time window (10 $\mu$sec) while for other REs, the magnetic order is quasi-static within the same time window, down to a temperature of 1.5 K.

# II. Experimental details

## A. Sample preparation and characterization

Polycrystalline $(Mo_{2/3}RE_{1/3})_2AlC$ samples are synthesized by pressureless sintering the following elemental powders: graphite (99.999 %), Mo (99.99 %) from Sigma-Aldrich, and Al (99.8 %) from Alfa Aesar, and RE (99.9 %) from Stanford Advanced Materials. Stoichiometric amounts of the elemental powders are manually mixed in an agate mortar, placed in an alumina crucible that is in turn inserted in an alumina tube furnace through which 5 sccm of Ar is flowing. The furnace is then heated, at 5 °C per minute, up to 1500 °C and held at 1500 °C for 10 h, before furnace cooling to RT. The loosely sintered powders are crushed into a fine powder that is directly used for further analysis.

$(Mo_{2/3}Gd_{1/3})_2AlC$ single crystals were grown using a flux growth technique in an induction-heated growth reactor and a sealed graphite crucible. Typical, initial elemental ratios $x$ before carbon incorporation were $x_{Mo} = 0.1$, $x_{Gd} = 0.4$ and $x_{Al} = 0.5$. After applying a maximum temperature of 1800 °C for 1 or 2 h, the growth reactor was slowly cooled down to 1000 °C in 2 to 7 days. The source of C is the graphite crucible walls, and typical C concentrations in the flux may range from $x_C = 0.2$-0.4, as estimated from weighting the crucible before and after growth. After growth, single crystals were extracted from the solidified flux by oxidizing the latter inside a chamber equipped with an air flux bubbling in water. Maximum lateral crystal size rarely exceeded one millimeter, and each growth resulted in hundreds of small single crystals with a wide distribution in size.

Room temperature powder X-ray diffraction (XRD) patterns of all compounds are recorded using the XPret Pro diffractometer using CuK$_\alpha$ radiation. A step size of 0.017° is used. The diffraction profiles are analyzed using Rietveld refinement as implemented within the FULLPROF suite.[9]

## B. Neutron powder diffraction

NPD of compounds with RE = Nd and Tb is performed at the D20 neutron diffractometer[10] located at the Institut Laue-Langevin (ILL), France. A wavelength of 2.4166 Å is obtained using a pyrolytic graphite monochromator (002 reflection) and no additional collimation. The samples are loaded into a vanadium holder of 8 mm diameter and cooled using a standard ILL Orange cryostat.[11]

High resolution NPD of RE = Er is performed at the High Resolution Powder Diffractometer for Thermal Neutrons (HRPT)[12] located at the Paul Scherrer Institute, Switzerland. An incident wavelength of 1.886 Å is obtained using a focusing Ge(511) monochromator with no additional collimation. The sample is loaded into a vanadium holder of 12 mm in diameter, cooled using a standard ILL Orange cryostat and measured in the high intensity mode.

Additional high resolution NPD measurements are performed on compounds with RE = Er and Ho using the BT-1 diffractometer[13] at the National Center for Neutron Research located at the National Institute of Standards and Technology, USA. An incident wavelength of 2.077 Å is obtained using the Ge(311) monochromator and an in-pile collimation of 60'. The samples are loaded into a vanadium holder with a diameter of 9.2 mm. A $^3$He cryostat[14] is used for the RE = Er sample, while the RE = Ho sample is loaded into a low $T$ high power CCR.[15]

NPD patterns for all measured compounds are collected at temperatures that range from the available base temperature (BT) up to the magnetic transition temperature and into the paramagnetic (PM) phase. An additional measurement of RE = Er is performed at 0.3 K using the $^3$He cryostat. Rietveld analysis of the data is performed using FULLPROF[9] suite. Symmetry analysis and propagation vector search are handled by BASIREPS[16] and SARAh,[17] respectively. Visualization of all crystal and magnetic structures in this paper is done using VESTA.[18]

## C. Muon spin rotation

All $\mu$SR measurements are done at the General-Purpose Spectrometer[19] (GPS) at the Paul Scherrer Institute, Switzerland. Powder samples of $(Mo_{2/3}RE_{1/3})_2AlC$ are pressed into pellets of 10 mm diameter and 5 mm thickness, mounted using aluminized Mylar tape and cooled down in a He gas flow cryostat.[20] All samples are measured from a base temperature of 1.5 K up to their respective paramagnetic temperatures.

Single crystals of $(Mo_{2/3}Gd_{1/3})_2AlC$ with lateral sizes greater than 100 μm are selected from each growth and glued on a silver plate sample holder using GE varnish. The crystallites with thickness ranging from 10 μm to 100 μm grow preferentially along their basal planes and therefore the $c^*$ axis is perependicular to their largest surface. The crystals are arranged such that the $c^*$ axis is perpendicular to the sample holder plane, but are not oriented in the *a-b* plane. The resulting sample is disk shaped with a diameter of 10 mm and a thickness of ≈3 mm. The single crystal sample is measured at 1.5 K for different initial muon polarization directions $\zeta = 4.7°$, 11°, 19°, 28°, 38° and 53° using two sets of detectors: Up-Down and Forward-Backward. The muon polarization direction is determined by measuring the sample at 30 K with an applied transverse field of 30 G and fitting the data with a simple cosine function. The fitted phase in the Up-Down detectors determines the muon polarization direction. All measurements are recorded in a time window of 3 μs with a 98 ps time resolution. Temperature dependent measurements of the single crystal sample were performed at the same temperatures as the RE = Gd powder sample.

## D. Density functional theory calculations

DFT calculations are performed using the QUANTUM ESPRESSO suite of codes.[21] The generalized gradient approximation (GGA) as parameterized by Perdew–Burke–Ernzerhof (PBE)[22] is used to describe exchange correlation effects and a non-spin polarized approach is used. All calculations are performed on a $6 \times 8 \times 4$ k-point grid with a 70 Ry (950 meV) cutoff energy with the wavefunctions of the 4*f* electrons treated as core states. The initial atomic positions for all materials are taken from the low temperature NPD measurements. The muon is approximated by a hydrogen atom occupying interstitial positions in the lattice. The search for candidate muon sites is performed by splitting the asymmetric unit of the $C2/c$ unit cell into a $3 \times 3 \times 3$ grid, with the muon placed in one of the 27 starting positions. All atoms in the unit cell are then relaxed until forces are smaller than $10^{-3}$ Ry/Bohr radius (26 meV/Å) and energy differences smaller than $10^{-4}$ Ry (1.4 meV). The final energy is then used to determine the most likely muon site.

# III.   Results

## A.  X-ray powder diffraction

XRD patterns of all samples (Fig. 1) contain reflections consistent with a monoclinic ($C2/c$) unit cell having lattice parameters (LPs) $a \approx 9.7$ Å, $b \approx 5.6$ Å, c $\approx 14.2$ Å, and $\beta \approx 103°$, with the $b$ axis defined as the unique axis. Additional reflections belonging to impurity phases in each sample are identified by systematically testing all phases containing RE, Mo, Al, C, and O. The main impurities found are $Mo_2C$ ($P63/mmc$), $Mo_3Al_2C$ ($P4_132$), and $RE_2O_3$ ($Ia$-3). Structural models for the main phase and all impurities are taken from Ref. 5. The refinement then consists of the main monoclinic $(Mo_{2/3}RE_{1/3})_2AlC$ compound with all identified impurities, with the unit cell parameters and atomic positions of the $(Mo_{2/3}RE_{1/3})_2AlC$ refined. Refinement results of the XRD data are given in Table I.

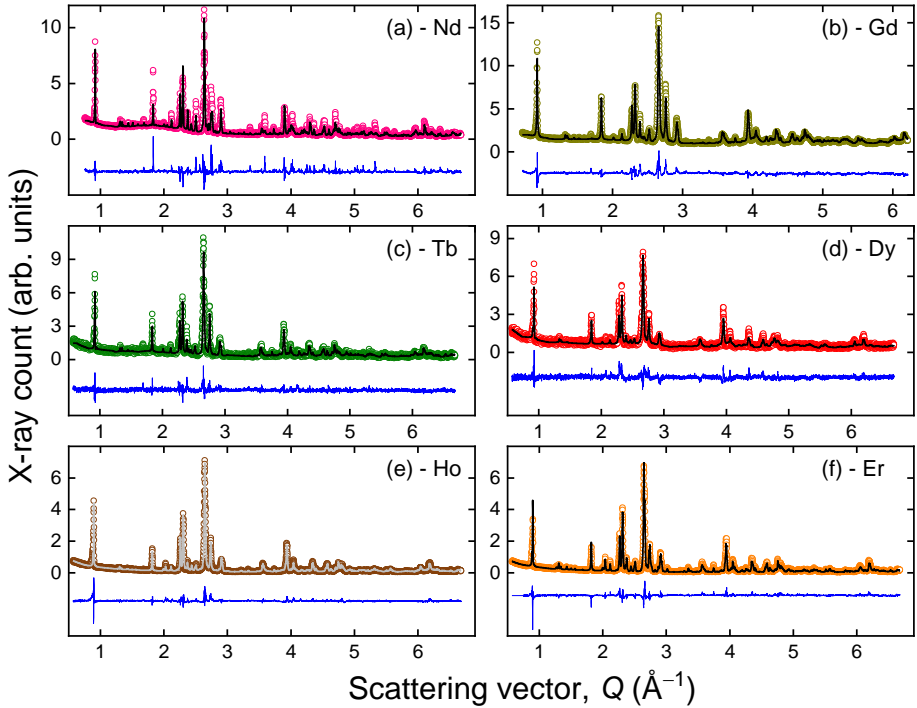

FIG. 1. Observed XRD patterns of $(Mo_{2/3}RE_{1/3})_2AlC$ compounds (symbols) obtained for RE = (a) Nd, (b) Gd, (c) Tb, (d) Dy, (e) Ho, and (e) Er at room temperature, the corresponding Rietveld refined profile (solid line on data), and their difference (solid bottom blue line). Uncertainties of data points ($1\sigma$) are represented by their spread around the refined profile.

Table I. LPs (*a*, *b*, *c*, *β*), unit cell volume (*V*), mass percent (*Wt*%), and agreement factor (*Rf*)[23] obtained from Rietveld refinement of room temperature XRD. Numbers in parentheses indicate statistical (1*σ*) uncertainties of the refinement.

| RE | *a* (Å) | *b* (Å) | *c* (Å) | *β* (°) | *V* (Å$^3$) | *Wt*% | *Rf* |
|----|---------|---------|---------|---------|------------|-------|------|
| Nd | 9.6675(3) | 5.5946(2) | 14.1700(4) | 103.481(2) | 745.27(4) | 89(1) | 17.6 |
| Gd | 9.5817(3) | 5.5440(2) | 14.1028(4) | 103.570(2) | 728.24(4) | 95.2(7) | 21 |
| Tb | 9.5420(4) | 5.5221(2) | 14.0601(6) | 103.541(2) | 720.26(5) | 91.6(9) | 16.7 |
| Dy | 9.5315(7) | 5.5059(4) | 14.0413(9) | 103.549(4) | 716.36(8) | 97(1) | 18.1 |
| Ho | 9.5075(3) | 5.5012(1) | 14.0306(4) | 103.527(2) | 713.48(3) | 93.1(8) | 16.5 |
| Er | 9.4876(2) | 5.4910(1) | 14.0130(3) | 103.524(1) | 709.79(3) | 90.1(5) | 7.47 |

## B. Ground state neutron powder diffraction

NPD patterns for all samples at the PM phase [*c.f.* Fig. 2(a)] contain reflections consistent with a monoclinic (*C2/c*) unit cell having LPs *a* ≈ 9.7 Å, *b* ≈ 5.6 Å, c ≈ 14.2 Å, and *β* ≈ 103°, as the main phase, in agreement with XRD measurements. For RE = Dy data and analysis, see Ref. 24. RE = Er data is examined in detail, with data of the magnetically ordered phase shown in Fig. 2(b). Upon cooling, clear additional reflections, originating from magnetic diffraction are observed and are emphasized by arrows in Fig. 2(b).

Refinement results from XRD (Fig. 1) are used as initial guesses for the paramagnetic NPD pattern [Fig. 2(a)]. Instrumental resolution and wavelength are obtained from measurement of standards and kept fixed. The refined variables are: LPs, atomic positions, scale factors, LPs of impurity phases, and instrumental zero shift. Following the refinement of the PM phase, a similar refinement is performed for BT with the additional magnetic phase excluded.

To isolate the magnetic contribution from the crystallographic reflections, the difference between BT and PM temperatures is calculated [Fig. 2(c) and Fig. S1 in the supplementary material] and analyzed first. To analyze the difference pattern between BT and PM temperatures, only the magnetic phase is considered. Its scale factor, LPs and atomic positions are fixed by the BT crystallographic refinement of the crystallographic reflections. Initial analysis of the difference pattern consists of a propagation vector search with simulated annealing based refinement, as implemented in SARAh.[17] To find possible spin configurations within the unit cell, irreducible representations for the little group of the $k$-vector are calculated using BASIREPS.[16] More details on the calculation and basis vectors is given in the supplementary material. For each magnetic representation, up to six basis vector coefficients can be refined. To find the best fitting magnetic configuration, a systematic refinement of all possible combinations of basis vectors is performed. The best choice is selected by requiring a minimum magnetic agreement factor ($R_{mag}$),[23] while also searching for the solution described by the least number of basis vector coefficients with physical magnetic moments. The final parameters for the BT magnetic structure are obtained by adding the magnetic contribution to the crystallographic profile and refining both magnetic and crystal structures simultaneously. In the final refinement, scale factors and LPs of all phases are refined, as well as atomic positions of the main phase.

Refinement of the BT magnetic structures reveals that the magnetic ground states of compounds with RE = Nd, Tb, Ho, and Er are SDWs with propagation vectors $\mathbf{k}_{RE}$ parallel to the crystallographic $b$ axis [Fig. 2(d)]. The magnetic moments of the RE atoms are aligned perpendicular to $\mathbf{k}_{RE}$ with magnetic directions varying with RE [Fig. 2(e)]. Three different magnetic configurations in the unit cell are identified and labeled using Bertaut notation[25] as $F_xF_z$ [Fig. 2(e), RE = Nd], $C_xC_z$ [Fig. 2(e), Tb, Er], and $C_xF_z$ [Fig. 2(e), Ho]. All three configurations are obtained from the $\Gamma_2$ representation (see supplementary material), where the magnetic moments of RE atoms 1 and 2 [Fig. 2(e)] are parallel and are oriented in the $a$-$c$ plane.

Attempting to refine the BT profile of RE = Er and Ho using the instrumental resolution, resulted in calculated reflections which are narrower than the observed reflections [Fig. 2(c), inset]. To obtain a better fitting peak shape, an additional Lorentzian contribution is added to the magnetic reflections (Fullprof $Y$ parameter[9]). The magnetic correlation length is then estimated using the Scherrer formula $\xi = 0.8\,\lambda/Y$.[26] For RE = Nd and Tb, no additional broadening is observed within the limits of the instrumental resolution, and thus $\xi$ cannot be determined, and assumed to match the crystallographic correlation length. RE = Nd shows the weakest reflections and has the worst agreement factor (Table II). To limit the number of refined parameters, the magnetic moments are constrained to lie in the $a$-$c$ plane in accordance with the other $(Mo_{2/3}RE_{1/3})_2AlC$ phases.

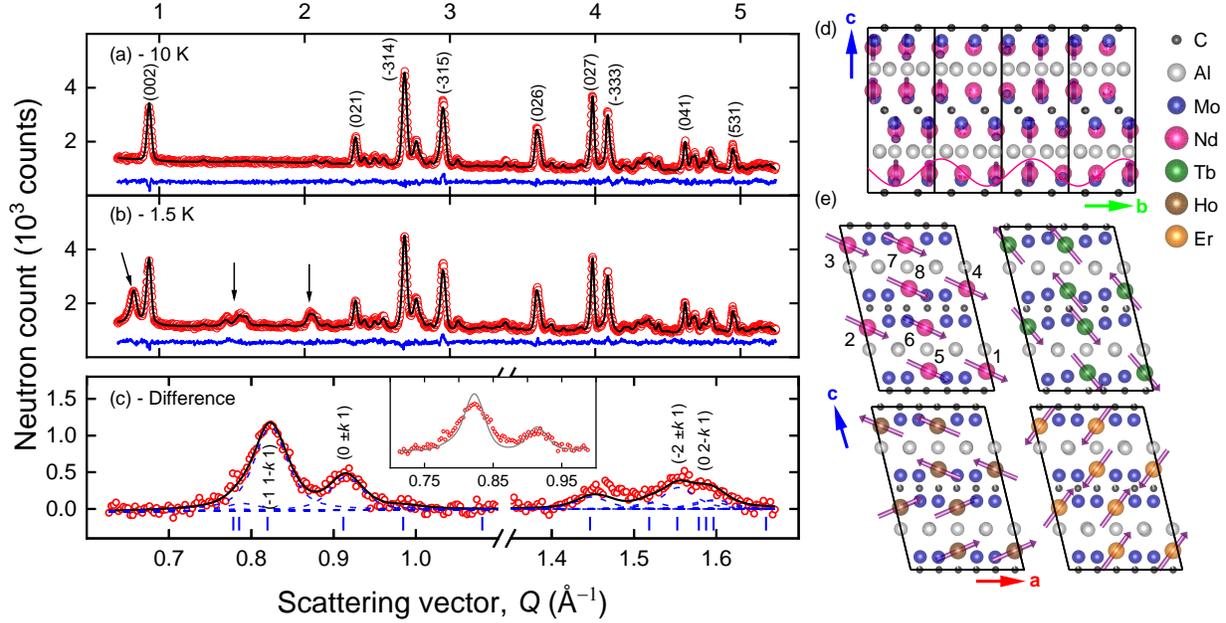

FIG. 2. Observed (symbols) neutron powder diffraction profiles of $(Mo_{2/3}Er_{1/3})_2AlC$ collected with the HRPT diffractometer at (a) 10 K, (b) 1.6 K, and (c) their difference. Arrows in (b) show magnetic reflections which appear upon cooling. The calculated profile and its difference from the data are shown as solid black and blue lines, respectively. Contributions from individual magnetic reflections in (c) are shown as dashed lines. Crystallographic and magnetic reflections of the main phase are labeled by their Miller indices and tick marks. Inset in (c) shows the calculated profile without including Lorentzian broadening (see text). Uncertainties of data points ($1\sigma$) are represented by their spread around the refined profile. (d) Depiction of the SDW structure for $(Mo_{2/3}Nd_{1/3})_2AlC$ in the $b$-$c$ plane. (e) Unit cell spin configurations of $(Mo_{2/3}RE_{1/3})_2AlC$ where RE = Nd, Tb, Ho, and Er at base temperature. The spin configuration for $(Mo_{2/3}Tb_{1/3})_2AlC$ shows the commensurate $\mathbf{k}_1 = (0, 0.5, 0)$ structure. The modulation in the magnetic moment magnitude acquired from the propagation vector (see text) is removed in this figure and all magnetic moments are rescaled for clarity. Numbers in (e) indicate order of magnetic atoms according to the international tables for crystallography.

Table II. Refined magnetic structure parameters for $(Mo_{2/3}RE_{1/3})_2AlC$ phases. $T_N$ is the Néel temperature, BT is the base temperature at which the measurement is performed, $m$ is the ordered magnetic moment per atom, $\theta$ is the orientation of the magnetic moment relative the $z$ axis [see Fig. 3], $k$ is the propagation vector component along the $b$ axis, $\xi_{RE}$ is the magnetic correlation length and $R_{mag}$ is the magnetic agreement factor.[9] Numbers in parenthesis indicate statistical uncertainty, while $\pm$ indicates systematic uncertainty.

| RE | $T_N$ (K)[5] | BT (K) | $m_{RE}$ ($\mu_B$/RE) | $\theta_{RE}$ (deg) | $k_{RE}$ | $\xi_{RE}$ (nm) | $R_{mag}$ |
|---|---|---|---|---|---|---|---|
| Nd | 7.6 | 1.5 | 1.76(2) | 115(2) | 0.730(2) | --- | 40 |
| Gd | 26 | --- | $6.6 \pm 0.3$ | $120 \pm 7$ | 0.5 | --- | --- |
| Tb | 20.1 | 3.5 | 5.63(3) | 132(2) | 0.5 | --- | 14.7 |
| | 28 | | 3.1(1) | 126(2) | 0.636(1) | --- | 29.2 |
| Ho | 7.8 | 4 | 8.7(2) | 67(1) | 0.6717(7) | 16(1) | 16.3 |
| Er | 3.6 | 1.6 | 6.54(6) | 36.6(9) | 0.6787(3) | 20.8(9) | 5.57 |

--- In case exists, falls below the limit of detection

Refinement of the BT measurement of RE = Tb requires a magnetic structure comprised of two propagation vectors (Table II). The basis vector coefficients' refinement is done for each propagation vector separately. The best fitting combination of basis vectors for both propagation vectors gives magnetic moments which are oriented in the $a$-$c$ plane. The final BT magnetic structures are shown in Fig. 2(d) and 2(e) and their properties are summarized in Table II.

## C. Ground state muon spin rotation

The asymmetry in the decay positrons at BT and zero applied field of different $(Mo_{2/3}RE_{1/3})_2AlC$ samples is plotted in Fig. 3(a) and 3(b). Three distinct behaviors are observed: a) RE = Nd shows oscillations which closely resemble a Bessel function. b) Similarly, RE = Gd shows oscillations which are well described by a cosine function. c) In contrast, other REs show an exponential decay which is characterized by a fast component ($t < 0.1\,\mu s$) and a slow component. For all compounds, the asymmetry signal decays to zero at long times ($\approx 3\,\mu s$, not shown) and does not show the characteristic 1/3 component of powders, which corresponds to a static magnetic field parallel to the muon spin.[27] Fourier transforms of the observed asymmetry in RE = Nd and Gd [Fig. 3(c) and 3(d)] show one dominant field in each sample, namely 0.35 T and 1.4 T for RE = Nd and Gd, respectively. The broadened field spectrum of RE = Nd is characteristic of an incommensurate magnetic structure,[28] while a single well-defined peak in the RE = Gd field spectrum corresponds to a simple commensurate structure [ferromagnet (FM) or AFM]. Following these observations, the asymmetry functions used to fit the data are:

$$A_1(t) = A_F J_0(\gamma_\mu B t + \phi) e^{-\lambda_F t} + A_S e^{-\lambda_S t} \tag{1}$$

$$A_2(t) = A_F \cos(\gamma_\mu B t + \phi) e^{-\lambda_F t} + A_S e^{-\lambda_S t} \tag{2}$$

$$A_3(t) = A_F e^{-\lambda_F t} + A_S e^{-\lambda_S t}, \tag{3}$$

where $F$ and $S$ stand for fast and slow relaxing signals, respectively, $A_{F/S}$ are the amplitudes of each term, $\lambda_{F/S}$ are the relaxation rates and $J_0$ is a zeroth-order Bessel function. $A_1$ is used for RE = Nd, $A_2$ for RE = Gd and $A_3$ is used for the rest. The final fit parameters for each RE (Table III) are obtained after simultaneous fitting of all temperature dependent spectra (see Sec. E).

Table III. Fit results for base temperature asymmetry of different $(Mo_{2/3}RE_{1/3})_2AlC$ compounds. $A_{F/S}$ is the fast/slow relaxing asymmetry amplitude, $B$ is the static magnetic field and $\lambda_{F/S}$ is the relaxation rate of the fast/slow component. Numbers in parenthesis indicate $1\sigma$ statistical uncertainties.

| RE | Asymmetry function | $A_F/A_S$ | $B$ (T) | $\lambda_F$ ($\mu s^{-1}$) | $\lambda_S$ ($\mu s^{-1}$) |
|---|---|---|---|---|---|
| Nd | $A_1$ | 3.29(2) | 0.350(2) | 14(1) | 0.42(1) |
| Gd | $A_2$ | 1.93(1) | 1.426(3) | 49(2) | 0.053(5) |
| Tb | | 4.43(5) | --- | 440(50) | 1.40(7) |
| Dy | $A_3$ | 4.53(7) | --- | 270(20) | 4.7(3) |
| Ho | | 9.2(2) | --- | 160(15) | 7.5(8) |
| Er | | 3.68(3) | --- | 113(4) | 3.2(1) |

--- Not relevant.

In order to convert the magnetic field distribution experienced by the muon into the magnetic moment distribution, the muon site has to be determined.[29] This is done in three steps. First, the stable muon site in RE = Nd is determined using an adaptation of DFT[30] which allows relaxation of the lattice in the presence of the muon. Second, the expected magnetic moment in RE = Nd is calculated using $\mu$SR results and compared to the NPD results to verify the validity of the DFT approach. Finally, the same procedure is applied to RE = Gd where the magnetic structure is not determined by NPD.

In RE = Nd, two candidate muon sites are obtained. They are labeled as A [Fig. 4(a)] with fractional coordinates (0.438, 0.111, 0.250) and B [Fig. 4(b)] with fractional coordinates (0.343, 0.114, 0.250). The two sites have nearly degenerate energy ($\approx 20$ meV difference). In contrast, for RE = Gd site A with coordinates (0.448, 0.087, 0.250) has the lowest energy by a margin of more than 100 meV. DFT calculations for other REs yield site A as the lowest energy site as well.

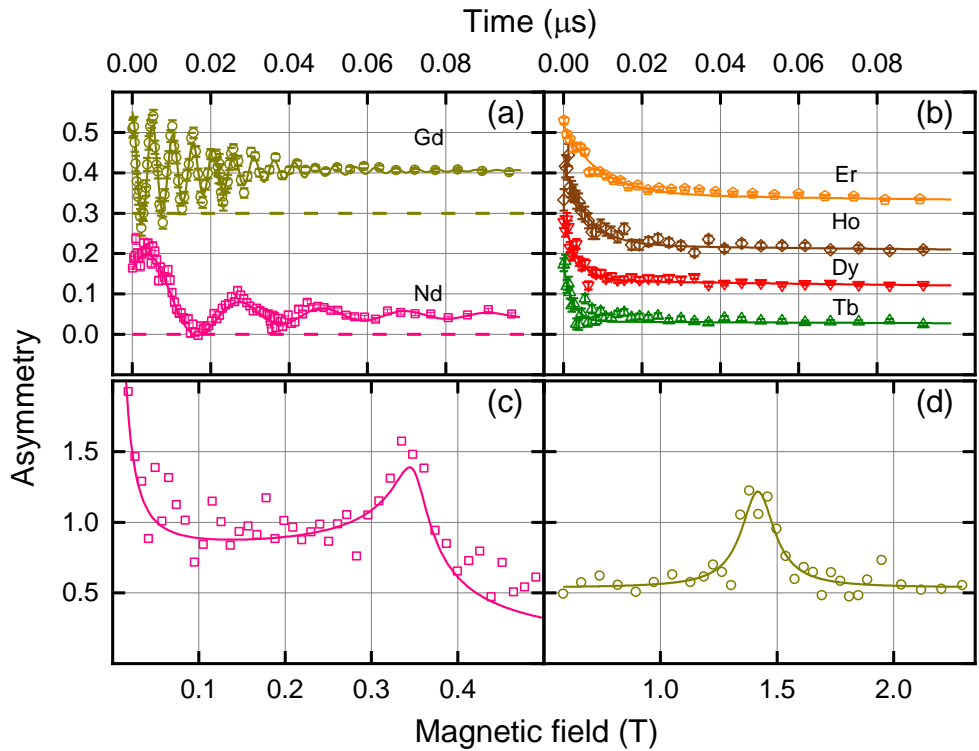

FIG. 3. Observed BT asymmetry (symbols) of (a) RE = Nd and RE = Gd, (b) RE = Tb, Dy, Ho, and Er. Data in (a) and (b) is shifted for clarity (for RE = Gd by 0.3 and for RE = Dy, Ho and Er by 0.1, 0.2, and 0.3, respectively). Error bars represent $1\sigma$ statistical uncertainty on the data points. Dashed lines in (a) indicate zero asymmetry for each data set. Magnetic field distribution for (c) RE = Nd and (d) RE = Gd obtained by Fourier transforming the data in (a). Solid lines show fits to the data.

After the determination of candidate muon sites in RE = Nd, the magnetic moment is calculated by assuming a given magnetic structure. The magnetic field at the muon site is determined by dipolar contributions from a $13 \times 13 \times 13$ unit cell grid centered on the muon site. Following the refinement strategy of the NPD data, four magnetic configurations, based on the basis vectors, namely $F_xF_z$ [Fig. 4(c)], $G_xG_z$, $C_xC_z$ [Fig. 4(d)], and $A_xA_z$, are considered. For each configuration, the moments are constrained to lie in the $a$-$c$ plane. A magnetic moment at site $i$ is then given by

$$\mathbf{m}_i = m_{SDW} \left( \sin\theta_{RE}\hat{\mathbf{x}} + \cos\theta_{RE}\hat{\mathbf{z}} \right) \cos\left( 2\pi\mathbf{k}_{RE}\cdot\mathbf{r}_i + \psi_{RE} \right), \tag{4}$$

where $m_{SDW}$ is the SDW amplitude, $\theta_{RE}$ is the magnetic moment angle relative to the Cartesian $z$ axis [Fig. 4(c)], $\mathbf{k}_{RE}$ is the propagation vector, $\mathbf{r}_i$ is the position of the $i$'th ion and $\psi_{RE}$ is a global magnetic phase. Since the magnetic structure of RE = Nd is incommensurate, $\psi_{Nd}$ is arbitrary and can be set to zero. A scan over $\theta_{Nd}$ is performed and $m_{SDW}$ corresponding to each $\theta_{Nd}$ is then calculated by matching the calculated magnetic field at the muon site to the observed field [Fig. 3(c)]. Among the resulting $m_{SDW}$-$\theta_{Nd}$ curves for muon site A [Fig. 4(e)], only the $F_xF_z$ configuration contains a ($m_{SDW}$, $\theta_{Nd}$) pair which matches the results obtained from NPD [Fig. 4(e), pink rectangle]. For muon site B, all calculated curves (Fig. S2) require $m_{SDW} > 2\,\mu_B$/Nd and thus do not intersect the NPD result. This indicates that even in the RE = Nd compound, site A seems to be the stable muon site and, it is concluded that site A is the most likely muon site candidate in all $(Mo_{2/3}RE_{1/3})_2AlC$.

Having confirmed the muon site and magnetic field calculations on RE = Nd, a similar process for determining possible magnetic structures in RE = Gd is performed. To limit the number of candidate magnetic structures, several constraints are placed on the magnetic moments in the RE = Gd unit cell. First, since there is only a single dominant frequency in the RE = Gd field spectrum, it can be safely assumed that the periodicity of its magnetic structure is commensurate with the crystal lattice. Because all $(Mo_{2/3}RE_{1/3})_2AlC$ phases in this study are isostructural, it is reasonable to assume that the propagation vector for RE = Gd has the same form as for the other compounds, i.e $(0, k_{Gd}, 0)$. Moreover, a single magnetic field requires $k_{Gd}$ to be either 0 or 1/2 as the magnetic moment magnitude must stay constant between adjacent unit cells.

Second, the possible spin configurations in the unit cell are limited to the four configurations tested for RE = Nd. These configurations ensure that atoms 1–4 in the unit cell [Fig. 2(e)] have the same magnetic moment, as required to produce a single magnetic field in all muon sites and are compatible with both $k_{Gd} = 0$ and $k_{Gd} = 1/2$ periodicities. Third, since the magnetic moments in compounds with RE = Nd, Tb, Ho and Er are oriented within the *a-c* plane, the same constraint is imposed on RE = Gd. The available fitting parameters are therefore $m_{SDW}$, $\theta_{Gd}$, and $\psi_{Gd}$.

Attempts to directly fit $m_{SDW}$, $\theta_{Gd}$ and $\psi_{Gd}$ by calculating the resulting asymmetry and minimizing the squared differences between calculated and observed values yield unstable results. Therefore, a brute force scan approach is employed. For $\mathbf{k}_{Gd} = (0, 0, 0)$, $\psi_{Gd}$ has no physical meaning and is set to zero. Scans of $\theta_{Gd}$ are performed for representations $\Gamma_1$, $\Gamma_3$, and $\Gamma_4$ (Table IV), since the FM representation $\Gamma_2$ can be ruled out from magnetization measurements.[5] Values for $\theta_{Gd}$, which are consistent with the magnetic field distribution of RE = Gd [Fig. 3(d)] are selected by requiring: (i) $m_{Gd} \leq 7\ \mu_B$, *i.e* lower than the $Gd^{3+}$ free ion moment.[31] (ii) $\Delta B \leq 0.05$ T, where the field inhomogeneity $\Delta B$ is the difference between the maximal and minimal field values among the muon sites and 0.05 T is the FWHM of the observed magnetic field distribution obtained by fitting it with a Lorentzian peak shape [Fig. 3(d), solid line]. Results consistent with conditions (i) and (ii) are found for $\Gamma_1$ and $\Gamma_3$ with Gd magnetic moment values ranging between 5.2 $\mu_B$ and 7 $\mu_B$. For $\mathbf{k}_{Gd} = (0, 0.5, 0)$ it is assumed that all magnetic moments in the $(Mo_{2/3}RE_{1/3})_2AlC$ unit cell have equal moments (i.e. $\psi_{Gd} = \pm 45°$) for simplicity. The physical moment on the Gd atom is denoted as $m_{Gd} = m_{SDW} \cos \psi_{Gd}$. Calculations for a general $\psi_{Gd}$ are treated in Sec. D. For each candidate structure, a scan over $\theta_{Gd}$ is performed in the range $0° \leq \theta_{Gd} \leq 180°$ [Fig. 4(f)]. For each $\theta_{Gd}$, the average magnetic field over eight muon sites in the unit cell $\bar{B}$ and $\Delta B$ [Fig. 4(g)] is computed. $m_{Gd}$ is calculated by requiring $\bar{B} = 1.426$ T, which is the observed average magnetic field (Table III). It is found that only configurations belonging to $\Gamma_2$ (Table IV) can satisfy conditions (i) and (ii) simultaneously and the resulting magnetic moment ranges between 6–7 $\mu_B$.

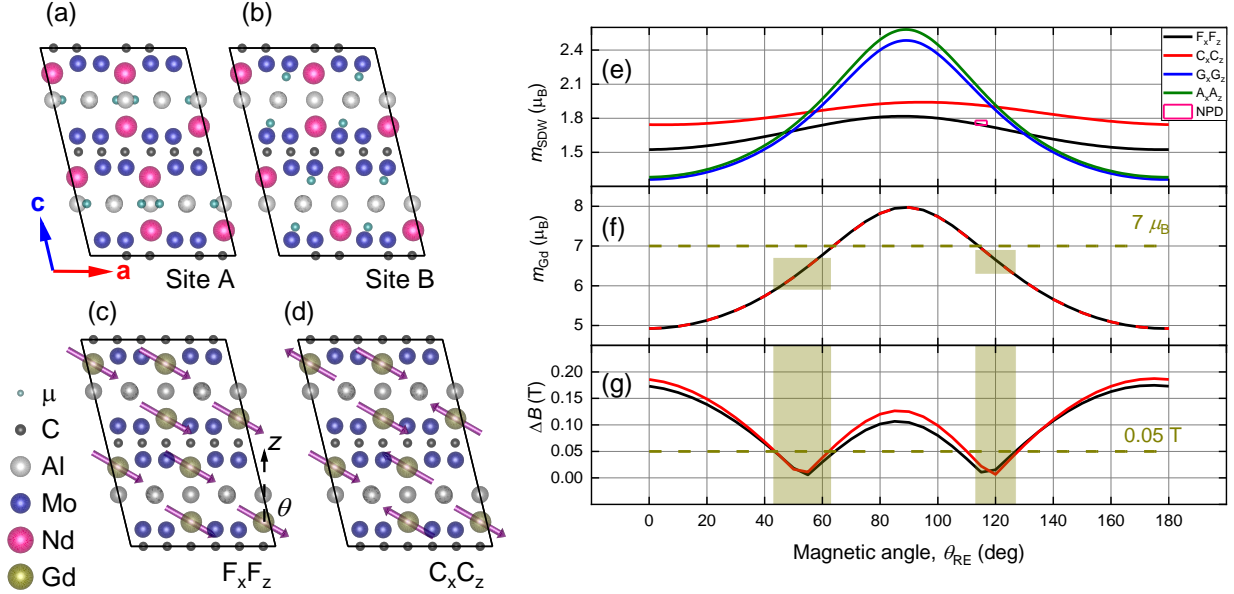

FIG. 4. Lowest energy muon (μ) site A (a) and muon site B (b) shown as dark cyan spheres. (c) The possible magnetic $F_xF_z$ configuration for RE = Gd in the *a-c* plane and the geometric definition of the magnetic angle *θ*. (d) The possible magnetic $C_xC_z$ configuration for RE = Gd. Both configurations in (c) and (d) have a propagation vector of $\mathbf{k}_{Gd} = (0, 0.5, 0)$. (e) Calculated SDW amplitude for RE = Nd as function of moments' direction $\theta_{Nd}$ for different spin configurations assuming muon site A. The NPD result (Table II) is given as a pink rectangle indicating uncertainties on the refined value. (f) Calculated physical magnetic moment for RE = Gd as a function of magnetic moment direction $\theta_{Gd}$ for $\mathbf{k}_{Gd} = (0, 0.5, 0)$ and assuming all magnetic moments in the unit cell have the same magnitude. (g) Calculated magnetic field inhomogeneity among muon sites in the unit cell for RE = Gd. Results consistent with observed *μ*SR data in (f) and (g) are highlighted using dark yellow rectangles.

Table IV. Scan results for possible magnetic configurations of $(Mo_{2/3}Gd_{1/3})_2AlC$ for two possible propagation vectors. The results are given as ranges for $\mathbf{k}_{Gd} = (0, 0, 0)$ and as bounds for $\mathbf{k}_{Gd} = (0, 0.5, 0)$. Parameters listed for modes of $\mathbf{k}_{Gd} = (0, 0.5, 0)$ are for the magnetic structures of highest symmetry (see text).

| Propagation vector, $\mathbf{k}_{Gd}$ | Representation | Mode | $m_{Gd}$ ($\mu_B$) | $\theta_{Gd}$ (deg) | $\psi_{Gd}$ (deg) |
|---|---|---|---|---|---|
| (0, 0, 0) | $\Gamma_1$ | $G_xG_z$ | $5.2 - 7$ | $0 - 48$ $130 - 180$ | 0 |
| | $\Gamma_2$ | $F_xF_z$* | | | |
| | $\Gamma_3$ | $A_xA_z$ | $5.2 - 7$ | $0 - 48$ $130 - 180$ | 0 |
| | $\Gamma_4$ | $C_xC_z$ | --- | --- | --- |
| (0, 0.5, 0) | $\Gamma_1$ | $G_xG_z$ | --- | --- | --- |
| | | $A_xA_z$ | --- | --- | --- |
| | $\Gamma_2$ | $F_xF_z$ | $6.3 \pm 0.4$ $6.6 \pm 0.3$ | $53 \pm 10$ $120 \pm 7$ | $-45^\dagger$ |
| | | $C_xC_z$ | $6.3 \pm 0.4$ $6.6 \pm 0.3$ | $53 \pm 9$ $121 \pm 7$ | $45^\dagger$ |

* Not considered.
$^\dagger$ Constrained to obtain equal moments on Gd atoms
--- No suitable pairs ($\theta_{Gd}$, $\psi_{Gd}$) found

## D. Muon spin rotation on $(Mo_{2/3}Gd_{1/3})_2AlC$ single crystals

Time dependence of the muon asymmetry is recorded using two perpendicular pairs of detectors. The first, forward - backward (FB) detectors measure the muon polarization along the muon momentum direction ($z$ axis). The second, up - down (UD) detectors measure the muon asymmetry perpendicular to the muon momentum ($x$ axis). The muon-site-averaged asymmetry $\overline{A}$ along each pair of detectors is calculated in appendix A with the results

$$\overline{A}_{UD}(t) = \tfrac{1}{4} A_{UD} \sin \zeta \, \tfrac{1}{8} \sum_{i=1}^{8} \left[ e^{-\lambda_f t} \cos\left( \gamma_\mu B_i t \right) \left( 3 + \cos 2\eta_i \right) + 2 e^{-\lambda_s t} \sin^2 \eta_i \right] + A_{UD}^{BG}(\zeta) \qquad (5)$$

$$\overline{A}_{\mathrm{FB}}\left(t\right) = A_{\mathrm{FB}} \cos \zeta \, \tfrac{1}{8} \sum_{i=1}^{8} \left[ e^{-\lambda_F t} \cos\left(\gamma_\mu B_i t\right) \sin^2 \eta_i + e^{-\lambda_S t} \cos^2 \eta_i \right] + A_{\mathrm{FB}}^{\mathrm{BG}}\left(\zeta\right), \qquad (6)$$

where $A_{\mathrm{UD}}$ and $A_{\mathrm{FB}}$ are the respective total detector asymmetries, $\zeta$ and $\eta$ are the angles of the initial muon spin and internal magnetic field relative the $z$ axis, respectively, and $A^{\mathrm{BG}}$ represents a background signal from muons stopped in the sample holder or in the glue between the crystallites. The index $i$ denotes the 8 muon sites in the $i$-MAX unit cell [Fig. 4(a)]. The magnetic field $B$ and angle $\eta$ are calculated from the magnetic structure parameters $m_{\mathrm{SDW}}$, $\theta_{\mathrm{Gd}}$, and $\psi_{\mathrm{Gd}}$ as a sum of dipolar fields.

BT asymmetry of the $(Mo_{2/3}Gd_{1/3})_2AlC$ single crystal sample is shown in Fig. 5, after background subtraction, for a number of muon spin angles $\zeta$. To find magnetic structures, which are consistent with the $\zeta$ dependence of the asymmetry, the powder data is first analyzed for the general $\psi_{\mathrm{Gd}}$ case. From this point on, we focus only on the case of $\mathbf{k}_{\mathrm{Gd}} = (0, 0.5, 0)$ as it is closer to the k-vectors of other RE's and is therefore more likely to be the correct option. Pairs of $(\theta_{\mathrm{Gd}}, \psi_{\mathrm{Gd}})$ are scanned in the ranges $0° \leq \theta_{\mathrm{Gd}} \leq 180°$ and $-90° \leq \psi_{\mathrm{Gd}} \leq 90°$ (Fig. 6). A slight change to condition (i) (see Sec. C) is required for the general case. The physical moments on the Gd atoms are given by $m_{\mathrm{SDW}} \cos \psi_{\mathrm{Gd}}$ and $m_{\mathrm{SDW}} \sin \psi_{\mathrm{Gd}}$ and condition (i) is changed to $\max(|m_{\mathrm{SDW}} \cos \psi_{\mathrm{Gd}}|, |m_{\mathrm{SDW}} \sin \psi_{\mathrm{Gd}}|) \leq 7 \, \mu_{\mathrm{B}}$. Only modes belonging to $\Gamma_2$ are consistent with (i) and (ii). For each mode, conditions (i) and (ii) are obeyed in four branches in the $\theta_{\mathrm{Gd}} - \psi_{\mathrm{Gd}}$ plane (Fig. 6). In general, the magnetic moment is found to vary between $5 \, \mu_{\mathrm{B}}$ and $7 \, \mu_{\mathrm{B}}$, $\theta_{\mathrm{Gd}}$ varies between $0°-60°$ or $110°-180°$, while $\psi_{\mathrm{Gd}}$ is centered around $-45°$ and $50°$ for $F_xF_z$, and around $-40°$ or $45°$ for $C_xC_z$.

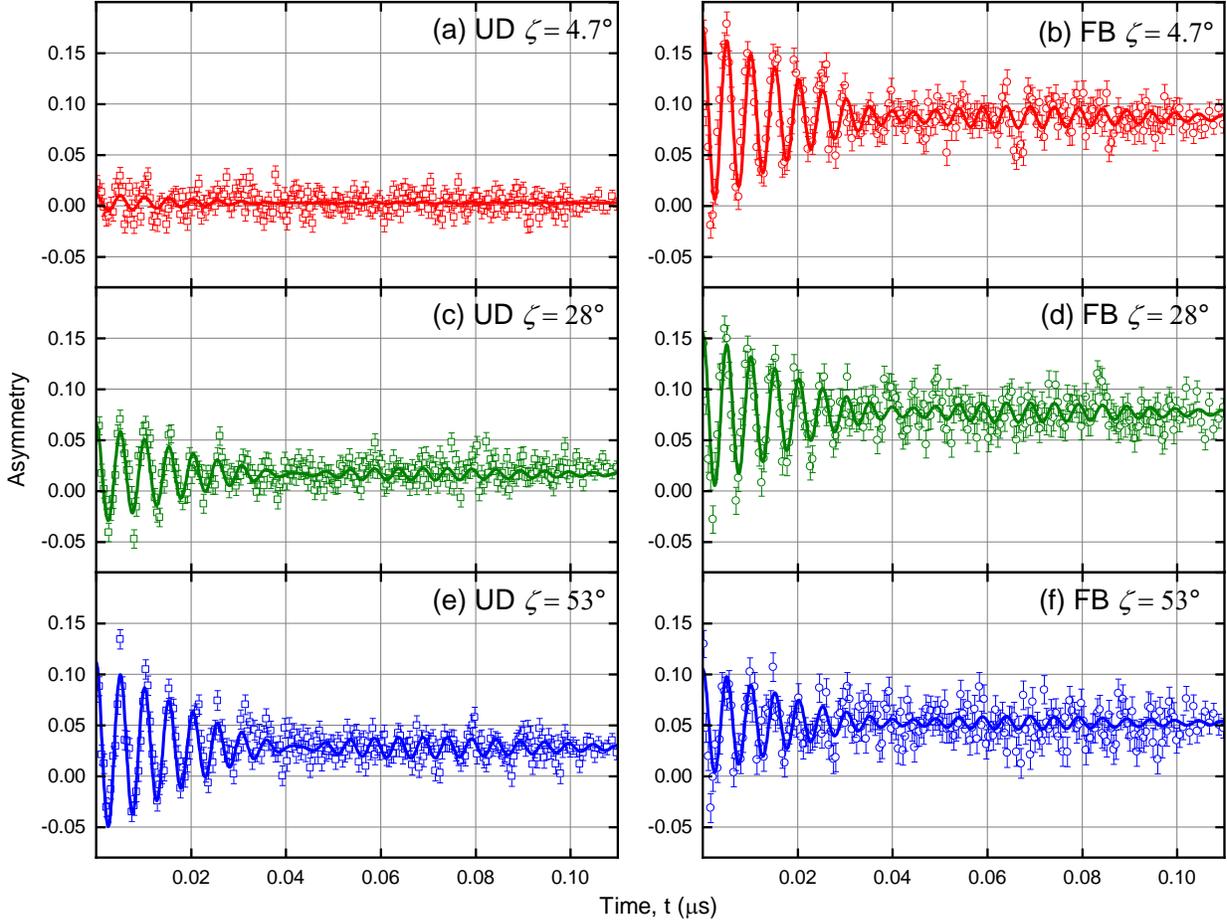

FIG. 5. Time dependence of the BT asymmetry (after background subtraction) for different initial muon spin angles $\zeta$. Left column shows the up-down detectors and right column – the forward-backward detectors. Solid lines are fits to equations (5) and (6). Error bars denote statistical ($1\sigma$) errors.

To fit the single crystal data, initial magnetic structure parameters are selected from one of the four branches in Fig. 6 as a strating guess for the fit. Equations (5) and (6) are fitted to the data sets of both detector pairs and all measured muon spin angles simultaneously. The fit parameters are constrained with conditions (i) and (ii) to stay consistent with the powder results. Different initial magnetic configurations from each branch are tested to check the stability of the fit. The final fit results are shown as 95 % confidence ellipses in Fig. 6 and solid lines in Fig. 5. Stable results are found in three out of the four branches for each magnetic configuration. The resulting magnetic moment for each fit result lies in the range 6.94–7 $\mu_B$ which matches the free ion moment of $Gd^{3+}$. The fitted values of $\psi_{Gd}$ in the branches that include $\psi_{Gd} = \pm 45°$ are found in the range $|\psi_{Gd}| = 45° \pm 1°$ for both magnetic configurations. This result agrees with our estimate for the most likely magnetic structure based on powder data alone, with the assumption that the magnetic moments in the unit cell have equal magnitudes (Table IV). Thus, these results put further constraints on the possible magnetic structures of $(Mo_{2/3}Gd_{1/3})_2AlC$.

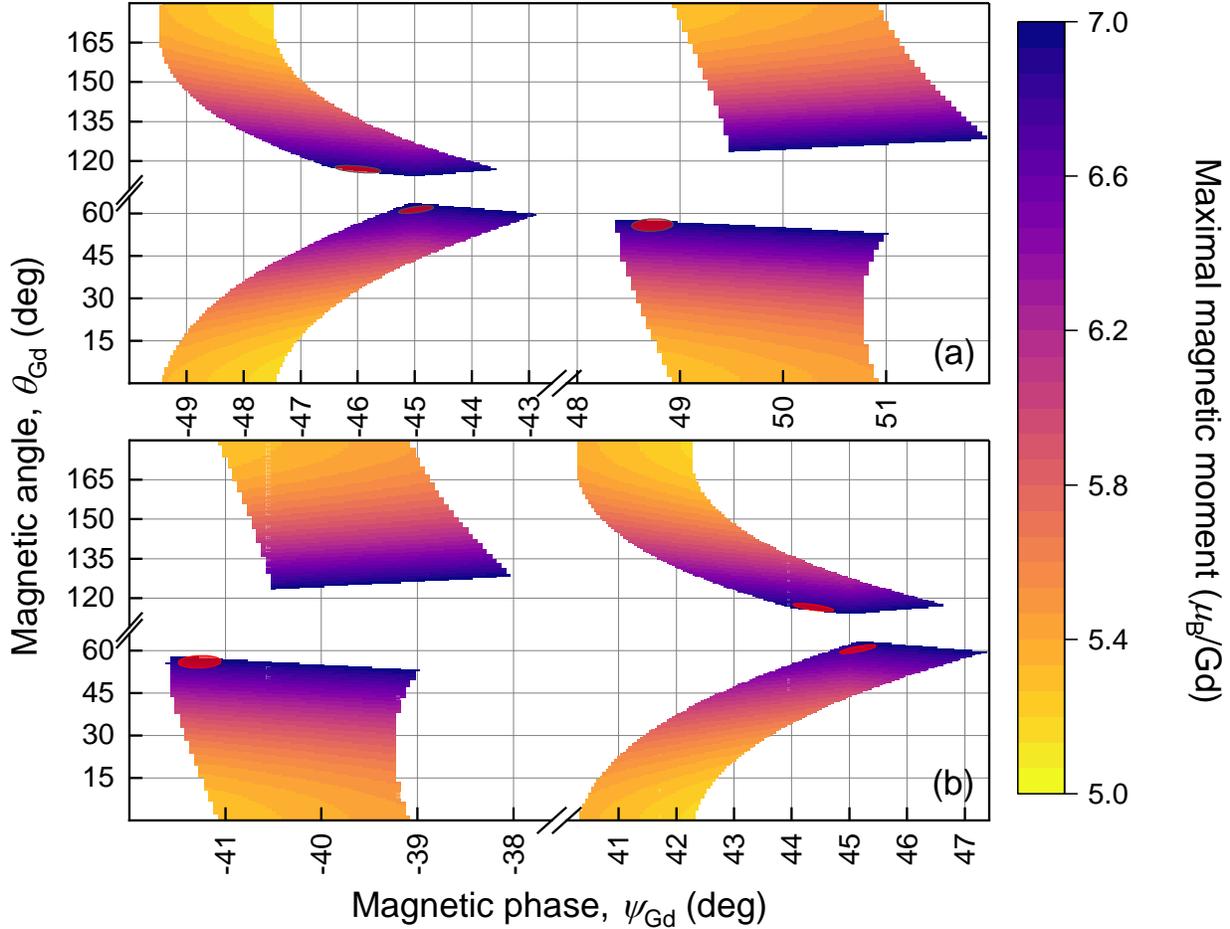

FIG. 6. Calculated maximal magnetic moment for $(Mo_{2/3}Gd_{1/3})_2AlC$ as function of $\psi_{Gd}$ and $\theta_{Gd}$. The calculations are performed assuming magnetic configuration (a) $F_xF_z$ and (b) $C_xC_z$. Colored areas show regions consistent with conditions (i) and (ii) that are required to reproduce the observed magnetic field distribution (see text). Red ellipses show 95 % confidence regions for magnetic structure parameters obtained by fitting the $\mu$SR data measured from RE = Gd single crystal sample (see text).

## E. Temperature evolution

### 1. Neutron diffraction

To obtain the temperature evolution of the magnetic and crystal structures of $(Mo_{2/3}RE_{1/3})_2AlC$ from the NPD data, a number of sequential refinements are performed. The initial refinement contains no constraints on the diffraction pattern parameters. The instrumental parameters (zero-shift and sample offset), scale factors and lattice constants of all phases, atomic positions of the main phase, and magnetic structure parameters are refined. After the initial refinement, the temperature evolution of each parameter is examined, and any parameter that shows no significant and systematic change with temperature is fixed to be the weighted average of all refined temperatures. The final refinement consists of the LPs, magnetic moment (magnitude and direction) and propagation vector refined, while all other parameters are fixed using their weighted averages. In the high-resolution NPD data of RE = Er and Ho, some atomic positions display a non-trivial temperature dependence and are refined as well.

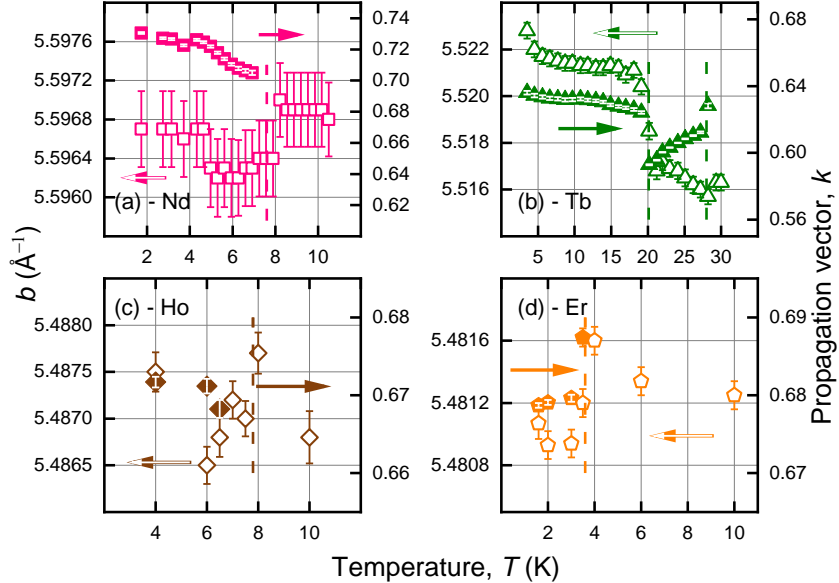

FIG. 7. Temperature evolution of the $b$ lattice parameter (open symbols, left ordinate) and propagation vector $k$ (full symbols, right ordinate) as determined by neutron diffraction. Dashed vertical lines indicate $T_N$ for each phase. Error bars in (b), (c) and (d) indicate statistical ($1\sigma$) uncertainties, while error bars in (a) contain a systematic uncertainty from the low fit quality for RE = Nd.

Temperature evolution of $\mathbf{k}_{RE}$ (Fig. 7) is observed for RE = Nd and Tb which is characteristic of incommensurate SDWs, while for RE = Ho and Er, $\mathbf{k}_{RE}$ stays constant with a value close to 2/3 indicating a commensurate SDW. The $b$ LP (Fig. 7) shows the strongest magnetostriction which is closely correlated with $T_N$ with up to an 0.08 % expansion for RE = Tb.

RE = Tb displays a magnetic structure which consists of two propagation vectors $\mathbf{k}_{Tb, 1} = (0, 0.5, 0)$ and $\mathbf{k}_{Tb, 2} = (0, 0.64, 0)$ (Table II). As $T$ increases, reflections belonging to $\mathbf{k}_{Tb, 1}$ decrease in intensity and show no relative intensity change between reflections. The direction of the $\mathbf{k}_{Tb, 1}$ magnetic moments is therefore kept fixed. Reflections belonging to $\mathbf{k}_{Tb, 2}$ do not show any significant changes up to 19 K, however both relative intensities and positions of the $\mathbf{k}_{Tb, 2}$ magnetic reflections change above that temperature. Above 20 K, reflections belonging to $\mathbf{k}_{Tb,1}$ disappear. The moment configuration of $\mathbf{k}_{Tb, 2}$ at these temperatures is identified by calculating the difference pattern and testing possible combinations of basis vectors. It is found that the magnetic structure belonging to $\mathbf{k}_{Tb, 2}$ changes from $C_xF_z$ to $C_xC_z$ (Fig. S3) above 20 K and therefore refinement consists of the $C_xC_z$ structure at these temperatures.

Temperature evolution of the refined magnetic moments of RE = Tb, Ho and Er are shown in Fig. 8(a). For RE = Tb, the magnetic moments of both propagation vectors reach saturation (upon cooling) below 15 K and switch magnitudes above 19 K. Slightly below $T_C$, the magnetic reflections in RE = Ho and Er become indistinguishable from the background due to strong diffuse magnetic scattering which results from short-range magnetic order in the sample. The abrupt drop in the magnetic moments of RE = Ho and Er in [upon heating, Fig. 8(a)] signifies the limit of detection of long-range magnetic order in the respective diffractometers.

The observed magnetic reflections of RE = Nd are too weak to be properly refined using the full crystallographic and magnetic patterns. The magnetic structure of this compound is therefore refined using the difference pattern for all temperatures and then kept fixed in the refinement of the full pattern. Due to the low fit quality of the RE = Nd magnetic structure, a more accurate temperature evolution of its magnetic moment [Fig. 8(b)] is obtained by taking the integrated intensity of the $(-1\ 1-k\ 4)$ reflection (Fig. S1). This reflection is chosen as it is strong and does not overlap with any nearby reflections. The magnetic moment of RE = Nd is then determined by normalizing the integrated intensity to the magnetic moment obtained from refinement at 1.5 K.

### 2. Muon spin rotation

Fitting of the temperature dependent muon asymmetry data of powder samples is first performed independently for all measured temperatures, including a fit of the detector efficiency ratio parameter $\alpha$. After examining the temperature behavior of the fitted parameters, a global fit is performed using equations (1), (2), and (3) with a global $\alpha$ parameter and a constraint $A_F + A_S = A_0$, with $A_0$ being a global fit parameter as well. For temperatures above which no oscillations are observed, $B$ and $\phi$ are constrained to be 0. ($Mo_{2/3}Gd_{1/3})_2AlC$ single crystal temperature dependent data is fitted to Eq. (7) with the constraint $A_F{}^D + A_S{}^D + A_{BG}{}^D = A_0{}^D$, where $A_0{}^D$ is a global fit parameter.

$$A(T,t) = A_F{}^D(T)\cos\left[\gamma_\mu B(T)t + \phi^D\right]e^{-\lambda_F(T)t} + A_S{}^D(T)e^{-\lambda_S(T)t} + A_{BG}{}^D e^{-\lambda_{BG}(T)t} \qquad (7)$$

Here $D$ denotes the pair of detectors UD or FB. The fit is first performed on all measured data sets individually. Then, a second fit is performed with parameters not depending on $T$ being shared between data sets, while parameters not depending on $D$ being shared between data sets of different detector pairs. Figure 8(b) presents the temperature evolution of the magnetic moment in the two magnets that show long time stability, namely RE = Nd and Gd. For clarity, the moment size of RE = Nd is multiplied by 3. The shaded regions represent confidence bands on the $\mu$SR moment size calculation. In RE = Nd good agreement is observed between NPD and $\mu$SR data. Clearly, RE = Gd is a stronger magnet in terms of moment size. When the temperature dependence of the single crystal Gd magnetic moment and relaxation rate [Fig. 8(b) and 8(c), full circles] is compared with the results from the powder sample [Fig. 8(b) and 8(c), open circles] good agreement is observed. This indicates that the magnetic structure of RE = Gd does not depend on the preparation method or sample type.

The temperature dependence of the magnetization of RE = Gd is analyzed by fitting an average moment $\langle S_z \rangle$ calculation, based on the Hamiltonian of a layered AFM

$$H = J\left(\sum_{i,\delta_\parallel} \boldsymbol{S}_i \cdot \boldsymbol{S}_{i+\delta_\parallel} + \alpha \sum_{i,\delta_\perp} \boldsymbol{S}_i \cdot \boldsymbol{S}_{i+\delta_\perp}\right) \qquad (8)$$

using Schwinger Boson mean field theory (SBMFT),[32–35] to the data. More details can be found in appendix B. In Eq. (8), $\delta_i$ denotes in-plane nearest neighbors and $\delta_\perp$ denotes out-of-plane nearest neighbors. The fit parameters and their best values are: $\alpha = 10^{-8.7\pm0.7}$, $m_0 = 6.9 \pm 0.3 \ \mu_B/Gd$ and $T_N = 29 \pm 1$ K, where $m_0$ is the calculated moment at $T = 0$ and $T_N$ is the calculated Néel temperature. The magnetic exchange interaction strength is $J = 2.38 \pm 0.03$ meV. The best fit is presented in Fig. 8(b) by the dark yellow solid line. The fit function is weakly sensitive to $\alpha$ and only its order of magnitude can be determined. Two additional calculations for $\alpha = 10^{-3}$ and $10^{-5}$ are added to demonstrate that as anisotropy decreases, $m(T)$ becomes $T$ independent at low temperature, in contrast to the data. The smallness of $\alpha$ supports the 2D nature of the RE = Gd magnet. In contrast, the magnetic moment of RE = Tb shows no $T$ dependence below 15 K which is consistent with a 3D behavior. RE = Er resembles anisotropic behavior as in Fig. 8(b), while the relatively high BT for RE = Ho does not allow clear determination of the $T \to 0$ behavior.

The muon relaxation rates in RE = Nd and Gd in Fig. 8(c) show classical spin lattice relaxation peak at, or close to, $T_N$ where the spin fluctuations are slow enough and their amplitude is large enough to effectively relax the muon polarization. For heavier REs, however, the situation is drastically different [Fig. 8(d)]. The relaxation rate in RE = Tb is nearly 5 times larger than in RE = Gd and shows a monotonic increase with decreasing temperature. Similar trends can be observed in the remaining compounds as well. The fast relaxing component in the muon asymmetry indicates that the magnetic fluctuations slow down to within the $\mu$SR time frame window of 10 ps–1 μs.[36,37]

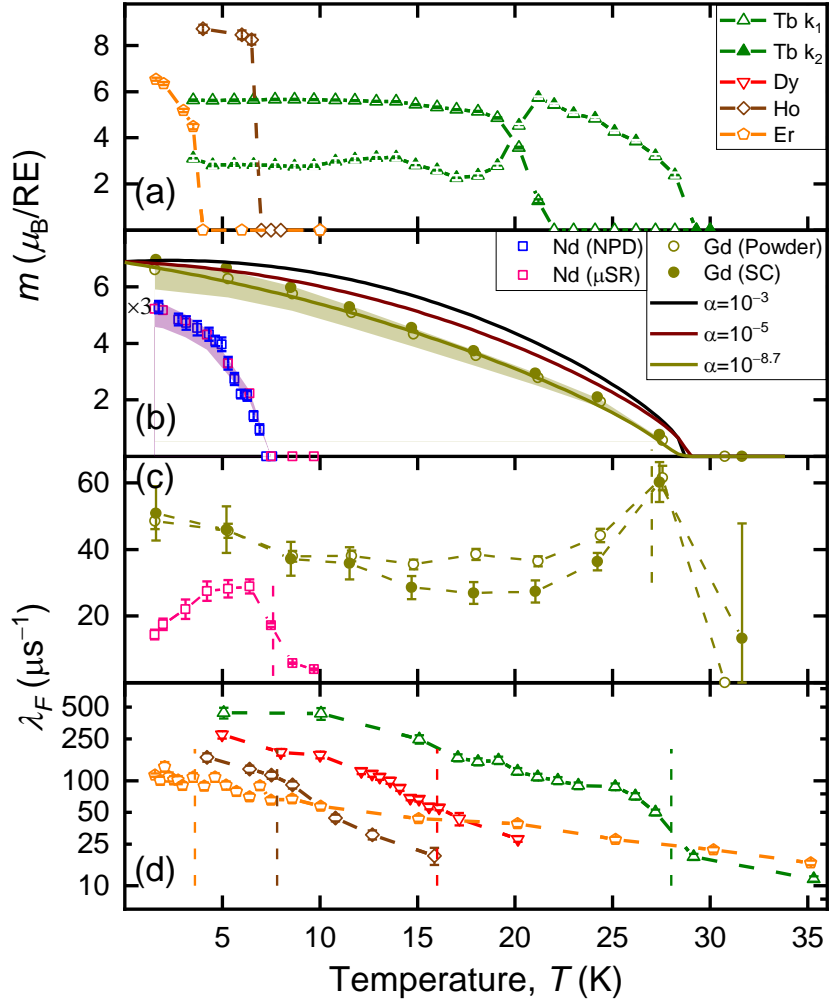

FIG. 8. Temperature evolution of the magnetic moment for $(Mo_{2/3}RE_{1/3})_2AlC$ with (a) RE = Tb, Ho and Er and (b) RE = Nd and RE = Gd. Pink squares in (b) are obtained from $\mu$SR by taking $\theta_{Nd} = 115°$ as measured by NPD (Table II). Open circles in (b) are calculated for $\theta_{Gd} = 120°$. The data for RE = Nd in (b) is multiplied by 3 for clarity. Solid lines show the calculated magnetic moment for RE = Gd for $\alpha = 10^{-3}$ (black line), $\alpha = 10^{-5}$ (dark red) and $\alpha = 10^{-8.7}$ (dark yellow, see text). Panels (c) and (d) show the temperature evolution of the muon relaxation rate for (c) RE = Nd and RE = Gd and (d) RE = Tb, Dy, Ho and Er. Dashed vertical lines in (c) and (d) show $T_N$ for each compound. Closed circles in (b) and (c) are obtained from fitting the single crystal data. Error bars on symbols represent $1\sigma$ statistical uncertainties, while systematic uncertainties in the magnetic moments obtained from $\mu$SR measurements are shown as confidence bands.

# IV. Discussion

The possible magnetic structures of RE = Gd [Fig. 4(c), (d)] are found by combining the NPD and $\mu$SR techniques. This is made possible due to the availability of isostructural compounds with different RE atoms, as well as the similarity in magnetic structures for different RE's and the relatively simple magnetic structure found for RE = Gd. While $\mu$SR can be used to refine subtle details in magnetic structures,[38,39] which cannot be easily determined from neutron measurements alone, the technique requires at least a rough magnetic structure to be known a priori. In this work, an application of $\mu$SR, which can predict a magnetic structure of a compound without any preliminary knowledge on the same compound, is demonstrated.

Temperature evolution of the RE = Gd magnetic moment [Fig. 8(b)] is well described by an SBMFT calculation up to and including $T_N$. This agreement indicates that critical fluctuations are either weak or occur in a very small temperature range around $T_N$ as SBMFT does not hold for strong fluctuations.[33] The obtained value for $\alpha$ is smaller than values found for some layered $CuO_2$ based materials[33] and is close to a calculated value of $2 \times 10^{-8}$ which was calculated for $Sr_2CuO_2Cl_2$ assuming magnetic dipole interaction between nearest neighbor layers.[33] This can indicate that interactions between planes in RE = Gd are also of dipolar origin.

When some magnetic properties are plotted against the RE atom size (Fig. 9), similar trends can be observed. $T_N/\langle T_N \rangle$, $\lambda_F/\langle \lambda_F \rangle$ and $m/m_{ion}$, measured at 1.5 K, where $\langle \rangle$ indicates average over RE, and $m_{ion}$ stands for the free ion moment, show a maximum for RE = Tb and follow an arc. The similarity in the RE dependence of $m/m_{ion}$ and $T_N/\langle T_N \rangle$ suggests that the reduction in the observed magnetic moment is due to temperature and that 1.5 K is not low enough to obtain the free ion moment for RE with $T_N < 10$ K. Indeed, the measurement of RE = Er at 0.3 K resulted in a magnetic moment of 7.5 $\mu_B$/Er bringing it closer to the free ion moment of 9 $\mu_B$.

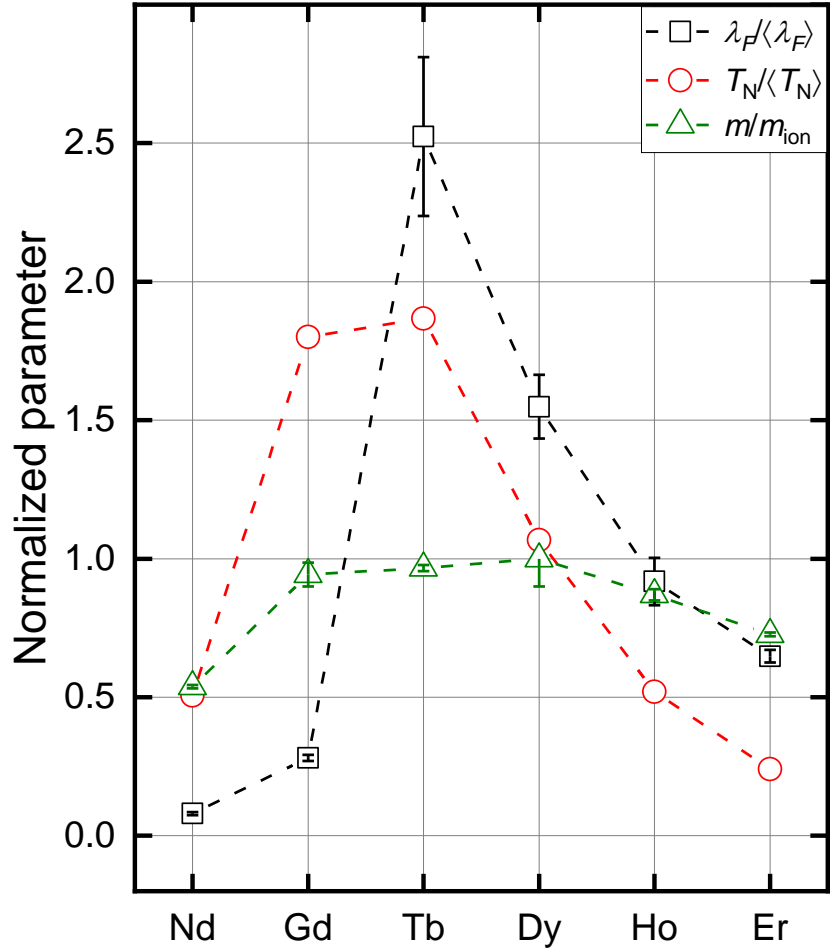

FIG. 9. Dependence of $\mu$SR relaxation rate $\lambda_F$, Néel temperature $T_N$, and magnetic moment $m$ measured at BT (4 K for $m$ of RE = Ho, and 1.5 K for the rest), on the RE element in $(Mo_{2/3}RE_{1/3})_2AlC$. For comparison, $\lambda_F$, and $T_N$ are normalized relative to their average over RE, while $m$ is normalized relative the free ion moment $m_{ion}$. Error bars for $\lambda_F$ and $m$ indicate $1\,\sigma$ statistical uncertainties. $T_N$ and $m$ data on RE = Dy is courtesy of Ref. 24.

While the muon relaxation observed for RE = Tb, Ho and Er may seem contradictory with NPD results at first, the observation of short-range ordering in RE = Ho and Er via broadening of the magnetic reflections [Fig. 2(c), inset] is evidence for a different nature in the magnetic ordering of the heavier lanthanides. The temperature dependence of the magnetic correlation length (Fig. S4) is similar to $\lambda_F$ and shows that the slowing down of the magnetic fluctuations is accompanied with a gradual increase in the size of ordered magnetic domains.

The nearly one to one correspondence between $\lambda_F$ and $T_N$ for REs heavier than Gd also indicates that the relaxation of the $\mu$SR signal in these compounds occurs due to a gradual freezing of the dynamic magnetic structure. As $T_N$ is lowered with increasing RE mass beyond Gd, the magnetic structure slows down less at 1.5 K resulting in shorter relaxation rates. The RE dependence of $T_N$, however, does not follow the well-known[40] de-Gennes factor given by $(g-1)^2 J(J+1)$, where $g$ is the Landé g-factor and $J$ is the total angular momentum of the RE ion, which has its maximum for RE = Gd. This suggests that the magnetic interactions in the $(Mo_{2/3}RE_{1/3})_2AlC$ phases are not described by RKKY interactions alone (as may be expected from an RE based magnet), but contain additional contributions, such as dipolar interactions between neighboring planes as mentioned for RE = Gd, which are also responsible for the emergence of spin fluctuations and short-range order in the heavy REs.

## V. Conclusion

From a crystal growth point of view, $(Mo_{2/3}RE_{1/3})_2AlC$ with RE = Gd, is the most promising compound for realization of a rare-earth based, high moment, 2D magnet out of the compounds studied here. However, as Gd is a strong neutron absorber, and the most informative experimental technique in magnetism, namely, NPD, is extremely challenging. Therefore, information from NPD on sister compounds was combined with information gained from $\mu$SR on all compounds, in order to determine for all variants, the most likely magnetic structures and their properties. $(Mo_{2/3}Gd_{1/3})_2AlC$ is found to be a stable commensurate AFM with the second highest $T_N = 29 \pm 1$ K of the series, a high magnetic moment of $6.5 \pm 0.5 \mu_B/Gd$, with one of two possible structures shown in Fig. 4(c) and 4(d), and magnetic anisotropy smaller than $10^{-8}$. This compound shows potential in providing new magnetic 2D derivatives for the field of Van der Waals heterostructures.

# Acknowledgements

The author is grateful to Prof. H. Shaked for discussions and help with analysis of the NPD data. A.P, A.K, D.P, O.R and E.N.C acknowledge the support of the Israel Atomic Energy Commission Pazy Foundation Grant. J. R. acknowledges support from the Knut and Alice Wallenberg (KAW) Foundation and the Flag-ERA JTC 2017 project "MORE-MXenes". Computational support is provided by the NegevHPC project (www.negevhpc.com). This work is partly based on experiments performed at the Swiss spallation neutron source SINQ, and the Swiss muon source (SμS), Paul Scherrer Institute, Villigen, Switzerland.

# Disclaimer

Certain commercial equipment, instruments, or materials (or suppliers, or software, ...) are identified in this paper to foster understanding. Such identification does not imply recommendation or endorsement by the National Institute of Standards and Technology, nor does it imply that the materials or equipment identified are necessarily the best available for the purpose.

# Appendix A

The geometry of the muon spin experiment is depicted in Fig. S5. The initial muon spin direction is rotated by an angle $\zeta$ from the $z$ axis in the $x$-$z$ plane

$$\mathbf{S}_0 = A_{\mathrm{UD}} \sin\zeta\, \hat{\mathbf{x}} + A_{\mathrm{FB}} \cos\zeta\, \hat{\mathbf{z}}, \tag{9}$$

while the direction of the magnetic field is denoted using standard spherical coordinates $(\eta, \chi)$

$$\mathbf{B} = B\sin\eta\cos\chi\, \hat{\mathbf{x}} + B\sin\eta\sin\chi\, \hat{\mathbf{y}} + B\cos\eta\, \hat{\mathbf{z}}. \tag{10}$$

Here, $A_{UD}$ and $A_{FB}$ denote the maximal asymmetry along each pair of detectors. The crystal flake is assumed to lie in the x-y plane. Time evolution of the muon spin, given an arbitrary magnetic field is given by

$$\mathbf{S}(t;\zeta,\eta,\chi) = \mathbf{S}_{\parallel}e^{-\lambda_s t} + \mathbf{S}_{\perp,1}e^{-\lambda_F t}\cos(\gamma|\mathbf{B}|t) + \mathbf{S}_{\perp,2}e^{-\lambda_F t}\sin(\gamma|\mathbf{B}|t), \tag{11}$$

where $\mathbf{S}_{\parallel} = (\mathbf{S}_0 \cdot \hat{\mathbf{B}})\hat{\mathbf{B}}$ is the component of the spin, parallel to the magnetic field and $\mathbf{S}_{\perp,1} = \mathbf{S}_0 - \mathbf{S}_{\parallel}$ and $\mathbf{S}_{\perp,2} = \hat{\mathbf{B}} \times \mathbf{S}_{\perp,1}$ are the two perpendicular components. $\lambda_F$ and $\lambda_S$ are the relaxation rates of the fast and slow decaying components, respectively. Since the flakes are randomly oriented in the plane, the total evolution of the muon polarization is obtained by averaging over $\chi$

$$\mathbf{S}_{avg}(t;\zeta,\eta) = \frac{1}{2\pi}\int_0^{2\pi}\mathbf{S}(t;\zeta,\eta,\chi)d\chi. \tag{12}$$

The polarization along each pair of detectors is obtained by projecting $\mathbf{S}_{avg}$ along the corresponding axes. By plugging equations (9), (10) and (11) into Eq. (12), we obtain

$$S_{UD}(t) = \mathbf{S}_{avg} \cdot \hat{\mathbf{x}} = \frac{1}{4}A_{UD}\sin\zeta\left[e^{-\lambda_F t}\cos(\gamma Bt)(3+\cos2\eta) + 2e^{-\lambda_s t}\sin^2\eta\right] \tag{13}$$

$$S_{FB}(t) = \mathbf{S}_{avg} \cdot \hat{\mathbf{z}} = A_{FB}\cos\zeta\left[e^{-\lambda_F t}\cos(\gamma Bt)\sin^2\eta + e^{-\lambda_s t}\cos^2\eta\right]. \tag{14}$$

Since there are 8 muon sites in the i-MAX unit cell, the final polarization is obtained by averaging over all muon sites, which gives equations (5) and (6).

# Appendix B

Calculation of the magnetic moment of RE = Gd is performed by using the nearly 2D antiferromagnetic Heisenberg model given in Equation (8). Details on the derivation of mean field equations for this Hamiltonian can be found elsewhere.[33–35] The self-consistent mean field equations which are solved are

$$\begin{cases} h = 2\alpha\left[S+1/2 - 2K(\Delta,h,t)\right] \\ K(\Delta,h,t) + K(\Delta,0,t) = S+1/2 \end{cases} \tag{15}$$

where $h$ is the staggered inter-layer mean field. $K$ is given by

$$K\left(\Delta,h,t\right)=2.32\int_0^1\frac{1+h+\Delta}{\omega\left(\Delta,h,\gamma\right)}\Big[n\left(\omega,t\right)+\tfrac{1}{2}\Big]\rho\left(\gamma\right)d\gamma \tag{16}$$

with the dispersion $\omega\left(\Delta,h,\gamma\right)=2.32\Big[\left(1+h+\Delta\right)^2-\gamma^2\Big]^{1/2}$, Bose occupancy function

$n\left(\omega,t\right)=\left(e^{\omega/t}-1\right)^{-1}$ and density of states $\rho\left(\gamma\right)=\frac{2}{\pi^2}\int_0^1\Big[\left(1-t^2\right)\left(1-t^2+\gamma^2t^2\right)\Big]^{-1/2}dt$ .

# Magnetic structure determination of rare-earth based, high moment, atomic laminates; potential parent materials for 2D magnets.

## Supplementary material

D. Potashnikov,[a,b] E. N. Caspi,[c] A. Pesach,[c] Q. Tao,[d] J. Rosen,[d] D. Sheptyakov,[e] H.A. Evans,[f] C. Ritter,[g] Z. Salman,[h] P. Bonfa,[i] T. Ouisse,[j] M. Barbier,[j] O. Rivin[c] and A. Keren[a]

[a] *Faculty of Physics, Technion - Israeli Institute of Technology, Haifa 32000, Israel.*

[b] *Israel Atomic Energy Commission, P.O. Box 7061, Tel-Aviv 61070, Israel.*

[c] *Department of Physics, Nuclear Research Centre-Negev, P.O. Box 9001, Beer Sheva 84190, Israel.*

[d] *Materials Design, Department of Physics, Chemistry, and Biology (IFM), Linköping University, SE-581 83 Linköping, Sweden.*

[e] *Laboratory for Neutron Scattering and Imaging, Paul Scherrer Institute, CH-5232 Villigen PSI, Switzerland.*

[f] *Center for Neutron Research, National Institute of Standards and Technology, Gaithersburg, MD, USA.*

[g] *Institut Laue-Langevin, 71 Avenue des Martyrs, 38000 Grenoble, France.*

[h] *Laboratory for Muon Spin Spectroscopy, Paul Scherrer Institute, CH-5232 Villigen PSI, Switzerland.*

[i] *Department of Mathematical, Physical and Computer Sciences, University of Parma, 43124 Parma, Italy.*

[j] *Univ. Grenoble Alpes, CNRS, Grenoble INP, LMGP, F-38000 Grenoble, France.*

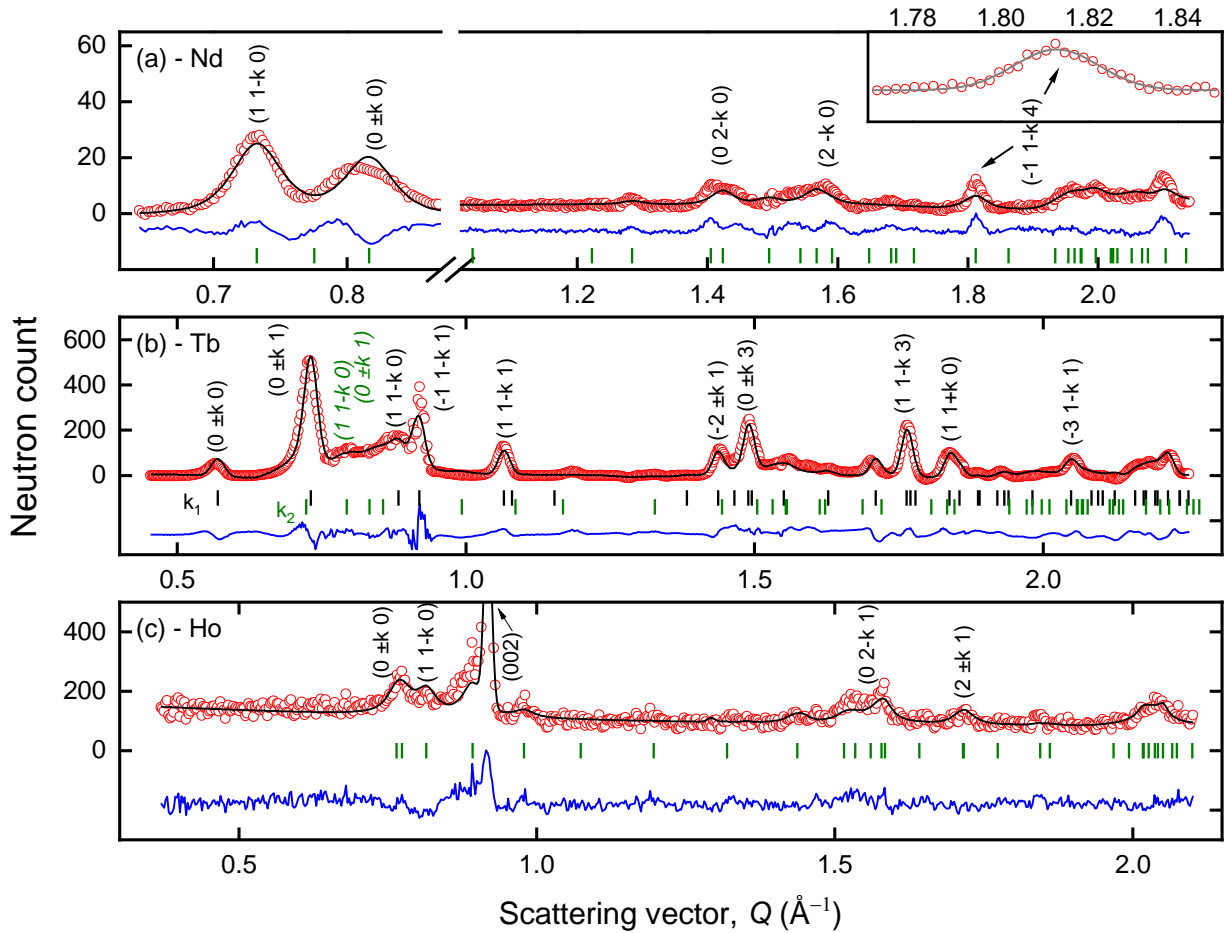

FIG. S1. Neutron count difference (symbols) between (a) 1.5 K and 10 K for RE = Nd and (b) 3.5 K and 30 K for RE = Tb obtained using the D20 diffractometer. (c) Observed neutron count for RE = Ho at 4 K obtained using the BT-1 diffractometer. The Rietveld refined profile is shown as a solid black line and its difference from the observed data is given by a solid blue line. Magnetic reflections are denoted by Miller indices and shown as tick marks. Inset in (a) zooms in on the (−1 1−*k* 4) reflection and shows a Gaussian fit used to obtain integrated intensity. RE = Tb contains reflections from both $\mathbf{k}_1$ and $\mathbf{k}_2$ (denoted using italics, and bottom tick marks). In (c) the crystallographic (002) reflection is also visible. Uncertainties of data points (1σ) are represented by their spread around the refined profile.

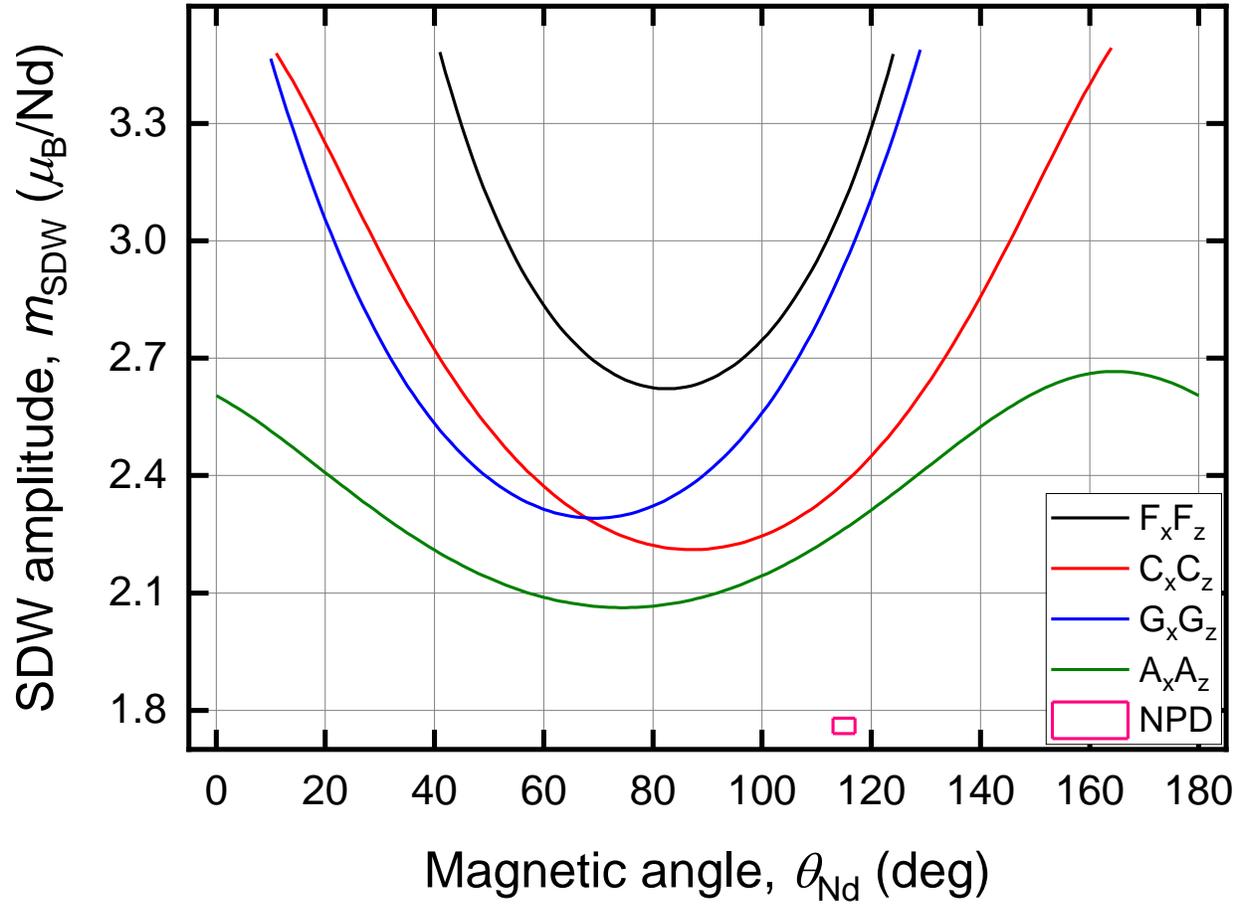

FIG. S2. Calculated SDW amplitude $m_{SDW}$ as function of magnetic angle $\theta_{Nd}$ for muon site B in $(Mo_{2/3}Nd_{1/3})_2AlC$. The pink rectangle shows bounds on $m_{SDW}$ and $\theta_{Nd}$ obtained from NPD.

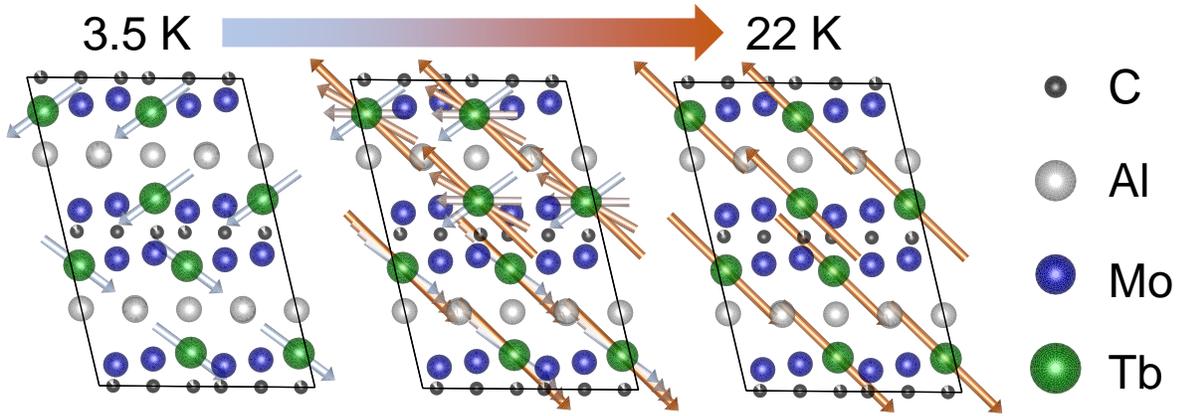

FIG. S3. Depiction of the transition of the $(Mo_{2/3}Tb_{1/3})_2AlC$ $\mathbf{k}_2$ structure from the $C_xF_z$ structure at 3.5 K to the $C_xC_z$ structure at 22 K. The color of the arrows represents different temperatures.

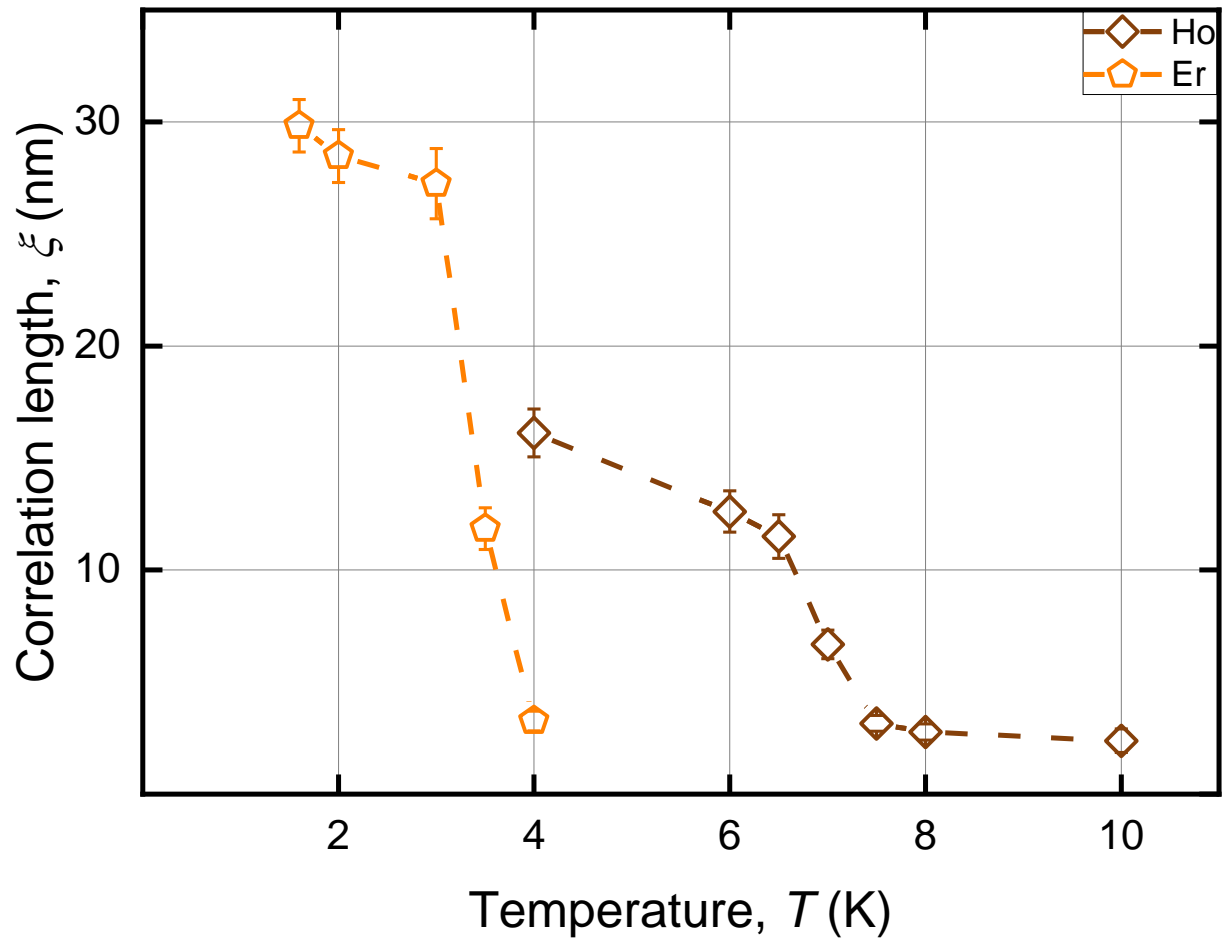

FIG. S4. Temperature dependence of the magnetic correlation length for $(Mo_{2/3}RE_{1/3})_2AlC$ with RE = Ho and Er as determined by high resolution NPD. Error bars indicate statistical ($1\sigma$) uncertainties.

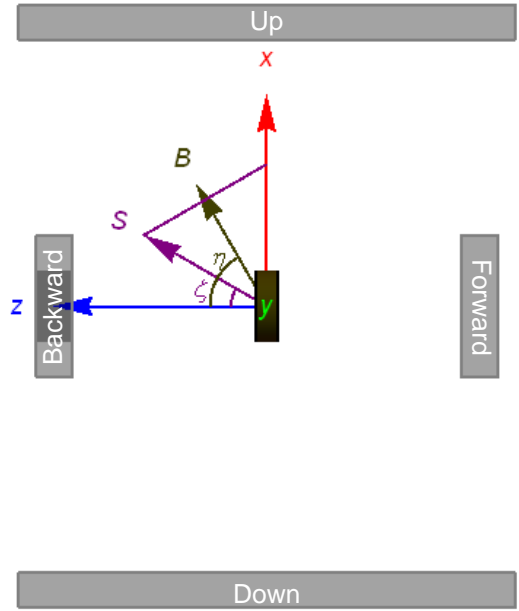

FIG. S5. Schematic representation of the detector geometry in the GPS muon spectrometer. The different pairs of detectors are shown as grey rectangles, the sample as a dark yellow disk. Arrows indicate the $x$ and $z$ axes with the $y$ axis pointing into the page. The magnetic field of the sample is shown as a dark yellow arrow and the direction of the muon spin as a purple arrow. The purple line shows the precession motion of the spin about the magnetic field.

**Representation analysis of the magnetic structure of (Mo$_{2/3}$RE$_{1/3}$)$_2$AlC**

To classify the possible magnetic structures, which are compatible with the crystal symmetry of (Mo$_{2/3}$RE$_{1/3}$)$_2$AlC, representation theory is used. Analysis of the magnetic representations is performed using the code BASIREPS. Here, an outline of the calculation steps is given and general notation to be used throughout this paper is defined. Since all propagation vectors are of the form $(0, k, 0)$ and do not occupy high symmetry points in the first Brillouin zone, the same analysis applies for all RE. The magnetic RE atoms occupy the 8$j$ site which consists of the following positions: 1. $(x, y, z)$, 2. $(1-x, y, -z + 0.5)$, 3. $(1-x, 1-y, 1-z)$, 4. $(x, 1-y, z+0.5)$, 5. $(x-0.5, y+0.5, z)$, 6. $(-x+1.5, y+0.5, -z+0.5)$, 7. $(-x+1.5, -y+0.5, 1-z)$, 8. $(x-0.5, -y+0.5, z+0.5)$. Atoms 5–8 are related to atoms 1–4 by a base translation $\mathbf{t} = (0.5, 0.5, 0)$ and therefore acquire a relative phase of

$$\psi_t = 2\pi\mathbf{k}\cdot\mathbf{t} \tag{17}$$

For atoms 5 and 6 $\psi_t = \pi k$ while for atoms 7 and 8 $\psi_t = -\pi k$. Therefore, only the relative phases between atoms 1–4 need to be considered. These atoms split into two orbits. Orbit I contains atoms 1 and 2, and orbit II contains atoms 3 and 4 with the relative phase between the orbits not determined by symmetry. For each orbit, there are two magnetic representations $\Gamma_1$ and $\Gamma_2$, each having 3 basis vectors (Table S1). Since the two orbits are independent, their basis vectors can be added and subtracted to obtain more symmetric representations as shown in Table S2. It is important to note that the new basis vectors still span the same space as the two independent orbits, and the relative phase between the two orbits is still a degree of freedom. For RE = Gd, magnetic structures with $\mathbf{k} = (0, 0, 0)$ are tested as well. In this case atoms 1–4 do not split into orbits and there are four representations each having three basis vectors, as shown in Table S3.

Table S1. Basis vectors of the irreducible representations for the little group of $\mathbf{k} = (0, k, 0)$.

| Representation | Atom / Basis vector | 1 | 2 |
|---|---|---|---|
| Γ₁ | V₁ | [1  0  0] | [–1  0  0] |
|  | V₂ | [0  1  0] | [ 0  1  0] |
|  | V₃ | [0  0  1] | [ 0  0 –1] |
| Γ₂ | V₁ | [1  0  0] | [ 1  0  0] |
|  | V₂ | [0  1  0] | [ 0 –1  0] |
|  | V₃ | [0  0  1] | [ 0  0  1] |

Table S2. Combinations of basis vectors of the irreducible representations of the little group of $\mathbf{k} = (0, 0.5, 0)$ for both orbits. The basis vectors are labeled using Bertraut notation.

| Representation | Atom / Basis vector | 1 | 2 | 3 | 4 |
|---|---|---|---|---|---|
| Γ₁ | $G_x$ | [1  0  0] | [–1  0  0] | [ 1  0  0] | [–1  0  0] |
|  | $F_y$ | [0  1  0] | [ 0  1  0] | [ 0  1  0] | [ 0  1  0] |
|  | $G_z$ | [0  0  1] | [ 0  0 –1] | [ 0  0  1] | [ 0  0 –1] |
|  | $A_x$ | [1  0  0] | [–1  0  0] | [–1  0  0] | [ 1  0  0] |
|  | $C_y$ | [0  1  0] | [ 0  1  0] | [ 0 –1  0] | [ 0 –1  0] |
|  | $A_z$ | [0  0  1] | [ 0  0 –1] | [ 0  0 –1] | [ 0  0  1] |
| Γ₂ | $F_x$ | [1  0  0] | [ 1  0  0] | [ 1  0  0] | [ 1  0  0] |
|  | $G_y$ | [0  1  0] | [ 0 –1  0] | [ 0  1  0] | [ 0 –1  0] |
|  | $F_z$ | [0  0  1] | [ 0  0  1] | [ 0  0  1] | [ 0  0  1] |
|  | $C_x$ | [1  0  0] | [ 1  0  0] | [–1  0  0] | [–1  0  0] |
|  | $A_y$ | [0  1  0] | [ 0 –1  0] | [ 0 –1  0] | [ 0  1  0] |
|  | $C_z$ | [0  0  1] | [ 0  0  1] | [ 0  0 –1] | [ 0  0 –1] |

Table S3. Irreducible representations of the little group of **k** = (0, 0, 0) and their basis vectors.

| Representation | Basis vector | Atom 1 | 2 | 3 | 4 |
|---|---|---|---|---|---|
| $\Gamma_1$ | $G_x$ | [1  0  0] | [−1  0  0] | [ 1  0  0] | [−1  0  0] |
| | $F_y$ | [0  1  0] | [ 0  1  0] | [ 0  1  0] | [ 0  1  0] |
| | $G_z$ | [0  0  1] | [ 0  0−1] | [ 0  0  1] | [ 0  0−1] |
| $\Gamma_2$ | $F_x$ | [1  0  0] | [ 1  0  0] | [ 1  0  0] | [ 1  0  0] |
| | $G_y$ | [0  1  0] | [ 0−1  0] | [ 0  1  0] | [ 0−1  0] |
| | $F_z$ | [0  0  1] | [ 0  0  1] | [ 0  0  1] | [ 0  0  1] |
| $\Gamma_3$ | $A_x$ | [1  0  0] | [−1  0  0] | [−1  0  0] | [ 1  0  0] |
| | $C_y$ | [0  1  0] | [ 0  1  0] | [ 0−1  0] | [ 0−1  0] |
| | $A_z$ | [0  0  1] | [ 0  0−1] | [ 0  0−1] | [ 0  0  1] |
| $\Gamma_4$ | $C_x$ | [1  0  0] | [ 1  0  0] | [−1  0  0] | [−1  0  0] |
| | $A_y$ | [0  1  0] | [ 0−1  0] | [ 0−1  0] | [ 0  1  0] |
| | $C_z$ | [0  0  1] | [ 0  0  1] | [ 0  0−1] | [ 0  0−1] |